\documentclass[iop]{emulateapj}
\usepackage{amsmath}

\usepackage{color}

\usepackage{lineno}
\shorttitle{Mergers by Evection Resonances}
\shortauthors{Bhaskar et al.}
\begin{document}

\title{Blackhole mergers through evection resonances}

\author{Hareesh Gautham Bhaskar \altaffilmark{1}, Gongjie Li \altaffilmark{1}, Douglas N. C. Lin \altaffilmark{2}}
\affil{$^1$ Center for Relativistic Astrophysics, School of Physics, Georgia Institute of Technology, Atlanta, GA 30332, USA}
\affil{$^2$ Department of Astronomy, University of California, Santa Cruz, CA, USA}


\begin{abstract}
Mechanisms have been proposed to enhance the merger rate of stellar mass black hole binaries, such as the Von Zeipel-Lidov-Kozai mechanism (vZLK). However, high inclinations are required in order to greatly excite the eccentricity and to reduce the merger time through vZLK. Here, we propose a novel pathway through which compact binaries could merge due to eccentricity increase  in general, including in a near coplanar configuration. Specifically, a compact binary migrating in an AGN disk could be captured in an evection resonance, when the precession rate of the binary equals their orbital period around the supermassive black hole. In our study we include precession to due first-order post Newtonian precession as well as that due to disk around one or both components of the binary. Eccentricity is excited when the binary sweeps through the resonance which happens only when it migrates on a timescale 10-100 times the libration timescale of the resonance. Libration timescale decreases as the mass of the disk increases. The eccentricity excitation of the binary can reduce the merger timescale by a factor up to $\sim 10^{3-5}$. 
%\li{It looks like $10^{3-4}$ in Fig. 9}. 

%We assume that a disk around one of the components of the binary as well as post-newtonian precession is responsible for the precession. Eccentricity excitation timescale depends on the migration timescale needed to ensure resonance sweeping. The migration timescale should always be less than the libration timescale of the resonance. We run also an ensemble of secular simulations to show the parameter space where the eccentricity can be excited using this mechanism. 
\end{abstract}

\keywords{ hierarchical triple systems --- secular dynamics --- evection resonance -- blackhole mergers}

\section{Introduction} 
Recent detections of gravitational waves has motivated investigations of mechanisms which allow compact objects to merge. Multiple merger pathways have been proposed in literature including isolated binary evolution \citep{dominik_double_2012,kinugawa_possible_2014,belczynski_first_2016,belczynski_origin_2018,giacobbo_merging_2018,spera_merging_2019,bavera_origin_2020}, gravitational capture \citep{oleary_gravitational_2009,kocsis_repeated_2012,gondan_eccentric_2018,zevin_eccentric_2019,rasskazov_rate_2019,samsing_single-single_2020} and dynamical evolution in clusters \citep{banerjee_stellar-mass_2017,oleary_binary_2006,samsing_formation_2014,rodriguez_binary_2016,askar_mocca-survey_2017,zevin_eccentric_2019,di_carlo_merging_2019}.
%and dynamical interactions in gas-rich nuclear regions \citep{mckernan_intermediate_2012,bartos_rapid_2017,stone_assisted_2017,mckernan_constraining_2018,leigh_rate_2018,yang_agn_2019,tagawa_spin_2020}.

In addition to above mentioned pathways, it has also been suggested that compact binaries could merge due to secular perturbations from distant perturbers. For instance, von Zeipel-Lidov-Kozai resonance \citep{von_zeipel_1910,lidov_evolution_1962,kozai_secular_1962} can excite the eccentricity of a compact binary to the point where gravitation radiation forces the binary to merge. This mechanism works more efficiently when the distant companion is highly inclined with respect to the binary (e.g., \cite{antonini_black_hole_2014}). However, it should be noted that the mutual inclination window where the mechanism is feasible can be substantially increased if the companion is on an eccentric orbit \citep{Li14_flip, hoang_black_2018} or if the companion is a binary instead of a single object \citep{liu_enhanced_2019}.  For near polar configurations it is also possible for the eccentricity to be unconstrained, allowing direct eccentric mergers \citep{antonini_black_2014,antognini_rapid_2014,Grishin18,fragione_black_2019}. This mechanism is shown to be viable in various environments including globular clusters \citep{antonini_black_2016,martinez_black_2020}, galactic nuclei \citep{petrovich_greatly_2017,antonini_secular_2012,hoang_black_2018,fragione_black_2019} and isolated triple systems\citep{silsbee_lidov-kozai_2017,antonini_binary_2017}.  

%It should be noted that this mechanism works only when the binary and the companion have a near polar initial configuration. 

Another source of gravitational waves could be AGN disks which are dense with stars and compact objects \citep{artymowicz1993,  mckernan_intermediate_2012,mckernan_constraining_2018,leigh_rate_2018,yang_agn_2019,tagawa_spin_2020}.  Multiple mechanisms can facilitate the merger of binaries in AGN disks. For instance,  gravitational torques from the disk and gas dynamical friction can shrink the binary orbit to a separation at which gravitational radiation forces a quick merger \citep{bartos_rapid_2017,tagawa_formation_2020}. In addition, binaries can also be hardened by flybys of other stars and compact objects in the disk \citep{leigh_rate_2018,stone_assisted_2017}.   

Merger rates for various merger pathways have also been estimated. For instance, due to uncertainty in the modelling of massive stars, the merger rate from isolated evolution is found to be in the broad range of $10^{-2}-10^3$ Gpc$^{-3}$ yr$^{-1}$ \citep{belczynski_first_2016}. Merger rate for secular three body interactions are estimated to be in the range $0.5-15$ Gpc$^{-3}$ yr$^{-1}$ \citep{antonini_binary_2017,petrovich_greatly_2017,hoang_black_2018}. In addition, mergers in clusters and galactic centers due to flybys are found to occur at a rate of $2-20$ Gpc$^{-3}$ yr$^{-1}$ \citep{fragione_black_2018}. Also, mergers in AGN disks due to stellar flybys and interactions with the gaseous disk are found to occur at a rate of $0.02-60$ Gpc$^{-3}$ yr$^{-1}$ \citep{tagawa_formation_2020}.  Meanwhile, estimates of merger rates from LIGO-Virgo observations are found to be in the range $15-38$ Gpc$^{-3}$ yr$^{-1}$ \citep{abbott_population_2021}. Given the higher than expected merger rate from observations as well as uncertainties in theoretically estimated merger rates, additional mechanisms may be needed to explain the observations.

In this paper we propose an alternative mechanism which involves binaries in an AGN disk under the influence of secular perturbations from the central super massive black hole. We show that the binaries migrating in an AGN disk can be captured in evection resonance which increases their eccentricity and reduces merger time. Evection resonance occurs when the precession rate of the binary equals the orbital period of the binary around a companion. It was originally used to show that the moon could have lost the excess angular momentum it acquired after impact with earth by exchanging it with the angular momentum of Sun-Earth orbit \citep{cuk_dynamical_2019, tian2017}. It has since been studied in the context of a wide range of astrophysical systems including evolution of natural satellites \citep{spalding_resonant_2016,cuk_secular_2004}, multiplanet systems \citep{touma_disruption_2015} and circumbinary planets \citep{xu_disruption_2016}. 

%Unlike the von-Ziepel-Lidov-Kozai resonance, capturing the binary in an evection resonance is not enough to sufficiently excite it's eccentricity \li{Please give a reference, otherwise it sounds like your result and the paragraph sounds like a "conclusion \& discussion" paragraph}. 

The eccentricity of a binary trapped in evection resonance can be excited if the components of the binary or the binary itself migrate in way which allows the binary to sweep through resonances. This is achieved by tidal migration for Earth-moon system and by disk migration for planetary systems. Migration rate determines the maximum eccentricity that can be achieved. If the migration rate is too fast, the system may leave the resonance impeding further eccentricity excitation. On the other hand if it is too slow, the maximum eccentricity is set by the timescale over which the migration mechanism operates. In this work, we assume that the compact binary is embedded in an AGN disk which would allow migration of both the binary and it's components.

\begin{figure*}
	\centering
	\includegraphics[height=0.3\linewidth]{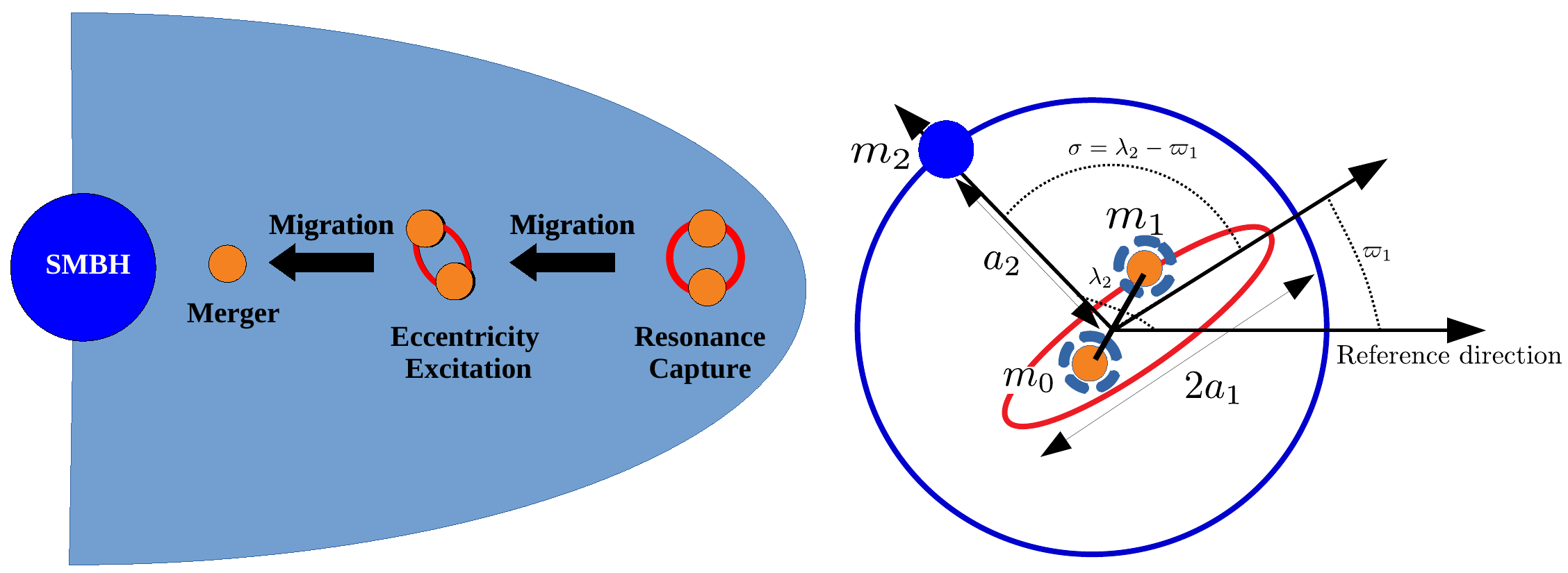}
	\caption{Schematic of the system. The basic mechanism discussed in this paper is shown on the left. A compact binary (shown as the orange dots) migrating in an AGN disk is trapped in evection resonance. As it migrates towards SMBH, its eccentricity increases and facilitates its merger. The configuration of the system is shown on the right, where the system contains a binary comprising of masses $m_0$ and $m_1$ on a Keplerian orbit with semi-major axis $a_1$, eccentricity $e_1$ and longitude of pericenter $\varpi$. The binary is being perturbed by a companion (SMBH) of mass $m_2$ on a circular orbit with semi-major axis $a_2$ and mean longitude of $\lambda_2$.$m_0$ and $m_1$ are surrounded by a disk (shown in blue) which contributes to a quadrupole interaction potential between the components of the binary. We assume that the system is hierarchical ($a_1 \ll a_2$) and the companion is much more massive as compared to the components of the binary ($m_2 \gg m_1,m_0$). 
	%{\color {red}  It may be useful to add in $\lambda_2$ and $\sigma$.} {\color {green} It may be useful to represent the disk with a dashed circle around $m_0$. Actually, it may be better to merge Figs 1 \& 2, i.e. side by side.}
	}
	\label{fig:cart_mech}
\end{figure*} 
%: for earth-moon system, it is the oblateness of earth, for circumbinary planets, it is the quadrupole potential of the binary stars and for multi-planetary systems it is mutual interactions between planets

Evection resonance also requires the orbit of the binary to precess. The source of precession depends on the system under consideration. In our study we consider two sources of precession: GR precession (to the first order post-newtonian correction) and precession due to a disk around components of the binary. 

This paper is organized as follows. In section \ref{section:analytical} we develop analytical theory for evection resonances. We discuss our numerical results in section \ref{sec:numres}. Our numerical integration scheme is detailed in subsection \ref{sec:numint}. Our disk migration model and merger timescales with evection resonance are discussed in subsections \ref{sec:secmig} and \ref{sec:mrgtsc}-\ref{sec:a1e1} respectively. Then, we discuss the resonance capture probability in section \ref{sec:prob}. Finally, we discuss our results and conclude in section \ref{section:conc}.

\label{section:intro} 
\section{Analytical Theory}
\label{section:analytical}

%\begin{figure}
%	\centering
%	%\includegraphics[height=0.8\linewidth,width=1.0\linewidth]{comp_traj.pdf}
%	\includegraphics[width=0.8\linewidth,height=0.6\linewidth]{schematic_fig.pdf}
%	\caption{Schematic of the system. The system contains a binary comprising of masses $m_0$ and $m_1$ on a keplerian orbit with semi-major axis $a_1$, eccentricity $e_1$ and longitude of pericenter $\varpi$. The binary is being perturbed by a companion of mass $m_2$ on a circular orbit with semi-major axis $a_2$ and mean longitude of $\lambda_2$ (not shown in the figure). We assume that the system is hierarchical ($a_1<<a_2$) and the companion is much more massive as compared to the components of the binary ($m_2>>m_1,m_0$). {\color {green} It may be useful to represent the disk with a dashed circle around $m_0$. Actually, it may be better to merge Figs 1 \& 2, i.e. side by side.} }
%	\label{fig:scheme}
%\end{figure}

%\subsection{Hamitonian formulation}

In this section we develop the Hamiltonian of the system near evection resonances. Figure \ref{fig:cart_mech} (Left) outlines the basic mechanism discussed in this paper. It shows a compact binary which is trapped in an evection resonance migrating in an AGN disk. As it migrates inward, the binary sweeps through resonances which increases its eccentricity, thereby decreasing its merger time. Figure \ref{fig:cart_mech} (Right) shows the schematic of the system in greater detail. It contains a binary comprised of objects with masses $m_0$ and $m_1$ in a Keplerian orbit with semi-major axis $a_1$, eccentricity $e_1$, and longitude of pericenter $\varpi_1$. $m_0$ and $m_1$ are surrounded by disks which contributes to a non-central interaction potential between the components of the binary. We model the interaction potential using a $J_2$ term similar to the potential of an oblate planet. The binary is being orbited by an object of mass $m_2$ (the SMBH) on a circular orbit with semi-major axis $a_2$ and mean longitude $\lambda_2$. Throughout this study, we assume that the system is in a hierarchical configuration ($a_2 \gg a_1$) and the companion is much more massive than either components of the binary ($m_2 \gg m_0,m_1$). In addition, the mutual inclination between the two orbits is taken to be zero.

The Hamiltonian of the system can be written as:
\begin{equation}
H=H_{J_2}+H_{m_2}+H_{GR} \label{eqn:hamiltonian}
\end{equation}
where $H_{J_2}$ is the term due to $J_2$ potential, $H_{GR}$ is due to first order post Newtonian correction and $H_{m_2}$ is due to the companion. For the purposes of this study, which is mainly concerned with secular changes in the orbit of the binary, it is useful to average the Hamiltonian over the mean anomaly of the binary. The averaged terms are given by:
\begin{eqnarray}
H_{J_2} &=& -\frac{GJ_{2,0}R_{d,0}^2 m_0 }{2a_1^3 (1-e_1^2)^{3/2}} -\frac{GJ_{2,1}R_{d,1}^2 m_1 }{2a_1^3 (1-e_1^2)^{3/2}} \nonumber \\
&=& -\frac{GJ_2R_d^2 m_0 }{2a_1^3 (1-e_1^2)^{3/2}}, \nonumber \\
H_{m_2} &=& -\frac{G m_2 a_1^2  \left(2+3 e_1^2+15 e_1^2 \cos (2 \lambda_2-2 \varpi_1)\right)}{8a_2^3}, \nonumber \\
H_{GR} &=& -\frac{3G^2(m_0+m_1)^2}{a_1^2c^2}\frac{1}{\sqrt{1-e_1^2}}, \nonumber
\end{eqnarray}
%$\mu_1=G(m_0+m_1)$, $\mu_2=G(m_0+m_1+m_2)$
where $G$ is the universal gravitational constant and $c$ is the speed of light (See \cite{murray_solar_1999,touma_resonances_1998,eggleton_orbital_2001}). We have defined an effective $J_2$ term given by:
\begin{equation}
    J_2 R_d^2 = \frac{m_0J_{2,0}R_{d,0}^2+m_1J_{2,1}R_{d,1}^2}{m_0}
\end{equation}
where $J_{2,0-1}$ are quadrupole moments of components of the binary,  $R_{d,0-1}$ are the radial extents of the disk. In the following we give a simple prescription which describes the $J_2$ potential in terms of properties of circum-blackhole disk.  Assuming a power law disk surface density(with index p) for circum-blackhole disks, we can write the quadrupole potential of the disk as:
\begin{equation}
    H_{J_2} = \frac{G}{4a_1^3(1-e_1)^{3/2}} \frac{2\pi\Sigma_{disc,0}}{(4-p)a_0^{-p}} \left( r^{4-p}_{out} - r^{4-p}_{in}\right) \label{eq:varj2def}
\end{equation}
where $r_{in}$ and $r_{out}$ are inner and outer radii of the disk and the surface density of the disk is given by $\Sigma_{disc} = \Sigma_{disc,0} (r/a_0)^{-p}$. In addition,  $\Sigma_{disc,0}$ is the characteristic density and $a_0$ is the characteristic length scale of the circum-blackhole disk. We assume that $r_{in} << r_{out}$. The outer extent of the disk is driven by the pericenter distance of the binary $r_{out} \approx a_1(1-e_1)/2$. As the pericenter distance of the binary decreases, the disk would be  truncated causing $J_2R_d^2$ to decrease. Hence, using Eqn. \ref{eq:varj2def} we can define $J_2R_d^2$ as a function of time (when $p < 4$):
\begin{equation}
    J_2(t) R^2_d(t) = J_2(0) R^2_d(0) \left(\frac{a_1(t)(1-e_1(t))}{a_1(0)(1-e_1(0))}\right)^{4-p} \label{eqn:varj2}
\end{equation}
In the following, we set $J_2 R_d^2$ to be constant to illustrate the dynamical effects, and in section \ref{sec:mrgtsc}-\ref{sec:prob} we set $p=0$ for a more realistic disk with constant density. In practice, we calculate $J_2$ at any point of time by using the minimum pericenter distance reached until that point in the evolution as shown in the above expression (see Figure \ref{fig:var_j2_p2}). In addition, we set $R_d=a_1(1-e_1)/2$ to get a reference on $J_2$ for equal mass binaries. We also compare the effects of different disk density profile with binary migration in Fig. \ref{fig:var_j2_p2}.
%and scales as: $J_2R_0^2 \propto m_{disk}R^2_{disk}/m_0$, where $m_{disk}$ is the mass of the disk and $R_{disk}$ is the radius of the disk.  The relevant length scale for disk size is the Laplace radius, $r_L=\sqrt[5]{J_2R^2_0a^3_1(m_0/m_1)}$, beyond which the disk is perturbed by $m_1$ and not coupled to the mass $m_0$. We can parametrize the disk mass using the following equation:
%\begin{equation}
 %   J_2R_0^2 = \frac{m_{disk}}{m_0}r_L^2 = \alpha_L r_L^2
%\end{equation}
%where $\alpha_L$ is the ratio of the mass of the disk with respect to the host object $m_0$. For equal mass binaries this ratio is independent of the mass of the components of the binaries. Table \ref{table:tabalpha} shows values of $\alpha_L$ for different values of $J_2R_0^2$ and $a_1$. We can see that for the values shown in the table, mass of the disk is a few percent of the mass of the host object.

It should be noted that in the above treatment we have only chosen terms corresponding to evection resonance and ignored other terms in the expansion of the Hamiltonian. For instance, when inner and outer orbits are not co-planar, secular terms corresponding to von Ziepel-Lidov-Kozai(vZLK) resonance should be included in the Hamiltonian\citep{cuk_secular_2004}. We have also ignored mean-motion resonances which can be important in less hierarchical systems (e.g., \cite{hamers_stability_2018}). In addition, eccentricity excitation due to capture in other resonances like eviction resonances has also been studied in literature but ignored in this study \citep{touma_resonances_1998,vaillant_eviction-like_2022}. Recently, it has been shown that the eccentricity of a binary in a hierarchical triple system  can be excited by precession induced resonances \citep{kuntz2022}. This resonance is important when the precession timescale
of the inner binary is equal to twice the period of the outer binary companion which is on an eccentric  orbit. In this study we assume that the outer companion is on a circular orbit and ignore such resonances.
%\begin{table}[]
%\caption{Values of $\alpha_L$ for different values of $J_2R_0^2$ and $a_1$.}
%centering
%\begin{tabular}{|l|l|l|}
%\hline
% &  $a_1 = 0.1 \text{AU}$                 & $a_1=1 \text{AU}$                   \\ \hline
%$J_2R_0^2=10^{-4} \text{AU}^2$                & $6.3\times 10^{-2}$ & $4.0\times 10^{-3}$ \\ \hline
%$J_2R_0^2=10^{-6} \text{AU}^2$                & $4.0\times 10^{-3}$ & $2.5\times 10^{-4}$ \\ \hline
%\end{tabular}
%\label{table:tabalpha}
%end{table}

We use a canonical transformation to the conjugate pair ($\sigma=\lambda_2-\varpi_1$,$\Sigma=\sqrt{G(m_0+m_1) a_1}(1-\sqrt{1-e^2})$) to simplify our analysis. The modified Hamiltonian can then be written as :
\begin{equation}
K = n_2\Sigma  + H(\Sigma,\sigma) \label{eq:kdef}
\end{equation}
where $n_2=\sqrt{G(m_0+m_1+m_2)/a_2^3}$ is the mean motion of the companion's orbit and $H(\Sigma,\sigma)$ is same as Eqn. \ref{eqn:hamiltonian} now written in terms of $\Sigma$ and $\sigma$. The time evolution of the system can then be obtained by solving Hamilton's equations:
\begin{eqnarray}
\dot{\Sigma} = -\frac{\partial K}{\partial \sigma} &\text{ , }& \dot{\sigma} =\frac{\partial K}{\partial \Sigma}. \label{eqn:eqcomp}
\end{eqnarray}

\begin{figure*}
	\centering
	\includegraphics[width=0.8\linewidth,height=0.6\linewidth]{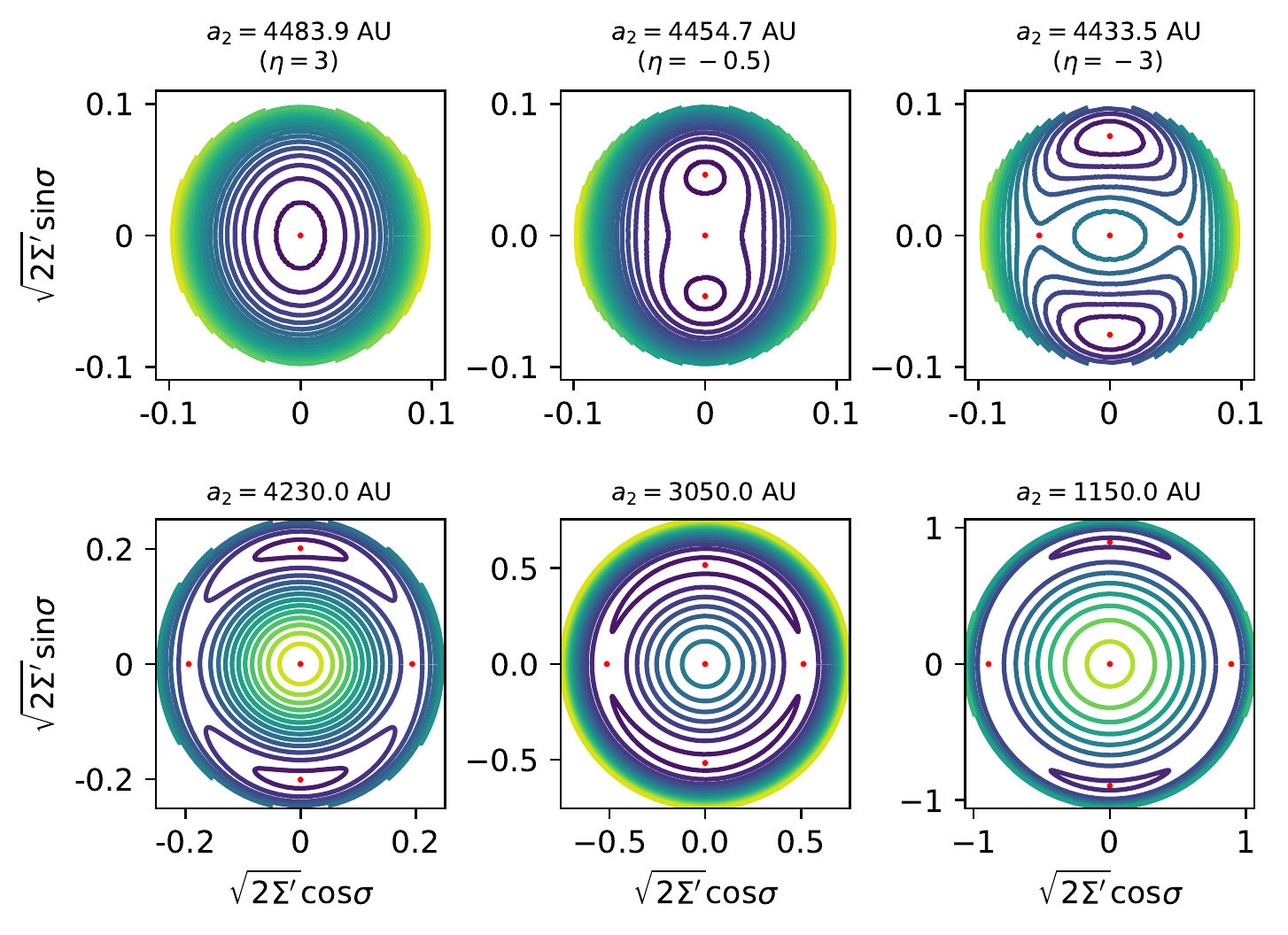}
	\caption{Contours of the Hamiltonian($K$). {\bf Top row:} The left panel corresponds to $\eta=3$ and has one stable fixed point at origin ($e_1=0$). The middle panel corresponds to $\eta=-0.5$ and has 3 fixed points. We can see that the stable point at the origin is unstable. The right most panel corresponds to $\eta=-3$ and has 5 fixed points. {\bf Bottom row:} Different panels correspond to different semi-major axis of the companion($a_2$). We can see that the eccentricity of the stable point increases as the semi-major axis of the companion decreases. In addition, the phase space structure of the hamitonian is preserved.  Fixed points at the origin and at $\sigma=\{\pi/2,3\pi/2\}$ are stable and fixed points at $\sigma=\{0,\pi\}$ are unstable. All fixed points are marked in red. It should be noted that the top row focuses on low eccentricities where $\sqrt{2\Sigma'} \approx e$. In this paper we focus on configurations with $\eta<1$ and assume that the system is trapped in a resonance at $\sigma=\pi/2$. We use the following parameters to make these plots: $a_1 = 1$AU, $J_2R_d^2 = 10^{-3}$AU$^2$ ($J_2=4 \times 10^{-3}$), $m_0 = 10M_\odot$, $m_2 = 10^6M_\odot$ and $m_1 = 10M_\odot$.}
	\label{fig:cont}
\end{figure*}

%The dynamics of this one degree of freedom system can be understood by looking at the contours of the hamiltonian. 
Using Eqn. \ref{eqn:hamiltonian} in Eqn. \ref{eq:kdef}, we can write the full Hamiltonian of the system as:
\begin{eqnarray}
K&=& n_2\sqrt{G(m_0+m_1)a_1}\Sigma' \nonumber \\ 
&&-\frac{3 G^2 (m_0+m_1)^2}{a_1^2 c^2(1-\Sigma')} -\frac{Gm_0J_2 R_d^2}{2a_1^3 \left(1-\Sigma' \right)^{3}} \nonumber \\
&& +\frac{a_1^2 G m_2 \left(15 (\Sigma' -2) \Sigma'  \cos (2 \sigma )+3 \Sigma'^2-6 \Sigma' -2\right)}{8 a_2^3} \nonumber \\ \label{eq:hamiltonianfull}
\end{eqnarray}
Where $\Sigma'= \Sigma/\sqrt{G(m_0+m_1)a_1}=1-\sqrt{1-e_1^2}$. The equations of motion are:
\begin{eqnarray}
\dot{\Sigma}'&=&-\frac{15 a_1^2 G m_2 (2-\Sigma') \Sigma'  \sin (2 \sigma )}{4a_2^3 \sqrt{a_1 G (m_0+m_1)}} \label{eq:Sigdot}\\
\dot{\sigma}&=& \frac{1}{\sqrt{a_1 G (m_0+m_1)}} \frac{3a_1^2 Gm_2 (\Sigma' -1) (5 \cos (2 \sigma )+1)}{4 a_2^3} \nonumber \\
&& - \frac{1}{\sqrt{a_1 G (m_0+m_1)}}\frac{3 G^2 (m_0+m_1)^2}{a_1^2 c^2 \left(1-\Sigma'\right)^2} \nonumber \\
&& -\frac{1}{\sqrt{a_1 G (m_0+m_1)}}\frac{3 G m_0 J_2R_d^2 }{2 a_1^3 \left(1-\Sigma'\right)^{4}}+n_2 \nonumber \\ \label{eq:sigdot} 
\end{eqnarray}

 The fixed points of the Hamiltonian can be derived by setting the above equations to zero i.e. $\dot{\Sigma}=\dot{\sigma}=0$. Using Eqn. \ref{eq:Sigdot}, we can see that $\dot{\Sigma}\propto \sin{2\sigma}$ which allows us to deduce that the fixed points of the Hamiltonian occur at $\sigma=\{0,\pi/2,\pi,3\pi/2\}$. Following the analysis of \cite{xu_disruption_2016}, we can expand the Hamiltonian (Eqn. \ref{eq:kdef}) near $e=0$ and rewrite it in a more compact form:
\begin{equation}
K_{\text{low-e}}=\eta \Theta + \Theta^2 +\Theta \cos \theta \label{eq:klowe}
\end{equation}
where, 
\begin{eqnarray}
\Theta &=& \frac{4 a_2^3 \left(a_1 G (m_0+m_1)^2+c^2 m_0J_2 R_d^2\right)}{5 a_1^5 c^2 m_2} (1-\sqrt{1-e^2}),\nonumber \\
&&  \\
\theta &=& 2\sigma \text{ and } \\
\eta &=& \frac{1}{5} + \frac{4 a_2^3 G (m_0+m_1)^2}{5 a_1^4 c^2 m_2} + \frac{2 a_2^3 m_0 J_2 R_d^2}{5 a_1^5 m_2} -\frac{4 a_2^3 n_1 n_2}{15Gm_2}. 
\end{eqnarray}
Here, $n_1=\sqrt{G(m_0+m_1)/a_1^3}$ is the mean motion of the binary orbit. 

%When $\eta>>1$, the dynamics is dominated by precession either due to GR or $J_2$ term.  Similarly, when $\eta<<-1$, the dynamics is dominated by the orbital motion of the binary around  the companion.  
When $\eta \gg 1$, the precession timescale due to either GR or the $J_2$ term is much shorter than the orbital period of the companion.  Similarly, when $\eta \ll -1$, the orbital motion of the binary around  the companion is much shorter than the precession timescale. Consequently, evection resonance is suppressed in both regimes. In contrast, when $|\eta| \approx 1$, the precession timescale and the orbital period of the binary around the companion are comparable. In this regime the resonant term is important and the binary can be captured in an evection resonance. 

The parameter $\eta$ also tells us about the number of fixed points in the system. The system always has a fixed at $e_1=0$ which is stable for $\eta >1$ and $\eta<-1$ and unstable elsewhere. For $\eta>1$, there is  only one fixed point i.e. the one at $e_1=0$. For $-1\leq\eta\leq1$, there are three fixed points and for $\eta<-1$, there are five fixed points.  Please note that the analysis of \cite{xu_disruption_2016} is valid only for low eccentricities. As we assume that the binary is captured in resonance at low eccentricities, it is justified to use the above approach in this work.

\begin{figure*}
	\centering
	\includegraphics[width=0.65\linewidth,height=0.35\linewidth]{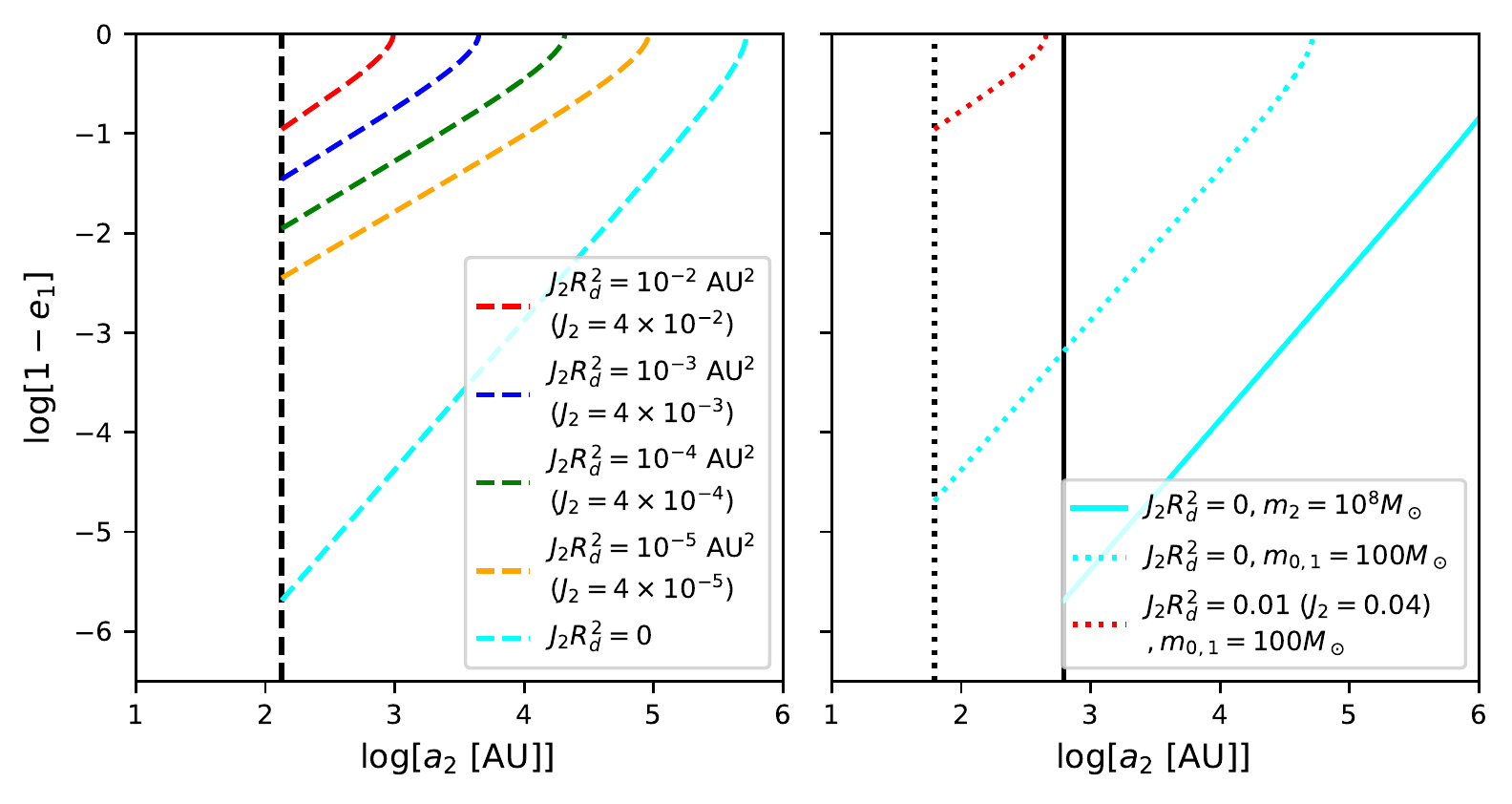}
	\caption{Location of stable points at $\sigma=\pi/2$. The left panel shows eccentricity at stable points as a function of $a_2$. The semi-major axis of the binary ($a_1$) is kept constant in this calculation at 1 AU.   We can see that once trapped in evection resonance, the eccentricity of the binary can be increased by decreasing $a_2$. We use the following parameters to make the plot in the left panel: $m_0 = 10M_\odot$,$m_1 = 10M_\odot$, $m_2 = 10^6M_\odot$  and $a_1=1$ AU.  
	In the right panel mass of the components of the binary and the companion are changed. Other parameters are kept constant. We can see that maximum eccentriciy decreases by increasing the mass of the binary and it remains same as the mass of the companion is increased. The black lines show $r_h/2$ (half of the Hill radius of the binary).}
	\label{fig:stbpts}
\end{figure*}

Figure \ref{fig:cont} shows contours of the Hamiltonian (Eqn. \ref{eq:hamiltonianfull}) for different values of $a_2$. In the top row,  the left panel corresponds to $\eta=3.0$, where the precession timescale is shorter than the orbital period of the companion.  %\li{Please remember not to focus on the math and be more physical}.
Consequently, we can see that there are no libration regions in this regime and that there is one fixed point at $e_1=0$. These contours resemble concentric circles and there is no eccentricity excitation in this regime. %\li{please remember to add the physical results} 
The middle panel of the top row corresponds to $\eta=-0.5$ and has three fixed points. While the fixed point at the origin is unstable, other two fixed points are stable. Thus, starting near origin with low eccentricity, the BH's eccentricity can be increased. Also, the right panel of the top row corresponds to $\eta=-3$ where the mean motion of the companion is faster than the precession rate of the binary, and the system has five fixed points. Fixed points at the origin as well as at $\sigma=\{\pi/2,3\pi/2\}$ are stable. The other two fixed points at $\sigma=\{0,\pi\}$ are unstable. Thus, starting with low eccentricity, the BH's eccentricity still cannot be excited. Contrarily, if the binary is captured in the libration region near $\pi/2$, the binary's eccentricity undergoes periodic oscillations. The bottom row shows contours of the Hamiltonian for systems in which the companion is closer to binary as compared to the top row. In this regime stable points occur at higher eccentricities  ($e>0.2$) and low eccentricity approximation is not valid (Eqn. \ref{eq:klowe}). The phase space structure is similar to the right panel in the top row and system has five fixed points.

In this paper we focus on the regime when $\eta\lesssim1$ where  the dynamics resembles the middle and the rightmost panel, which corresponds to the region where the binary orbital mean motion is comparable with the precession. We assume that the system is captured in the resonance at $\sigma=\pi/2$. This 
supposition is justified because starting from low eccentricity, a binary can always be trapped in the resonance as long as the migration timescale is sufficiently longer than the libration timescale (See Figure 7 of \cite{xu_migration_2016}). We note that our results are the same for resonances at $\sigma=3\pi/2$.

The libration regions correspond to evection resonances where the precession rate due to the $J_2$ and post-Newtonian terms equals the orbital period of $m_2$ around the binary. The exact location of these fixed points can be deduced by solving the following equation:
\begin{equation}
\dot{\sigma}(a_1,a_2,e_1)|_{\sigma = \{0,\pi,\pi/2,3\pi/2\}}=0. \label{eqn:sbeq}
\end{equation} 
These are shown as red points in each of the panels in Fig. \ref{fig:cont}. Using the above relation, we can show that near evection resonances ( $\sigma=\pi/2$) at low eccentricities ($e\rightarrow$0), $a_2 \propto a_1^{5/3}$ when GR precession dominates and $a_2 \propto a_1^{7/3}$ when $J_2$ precession dominates.
%While the above equation can be solved numerically, a simplified expression for the location of stable points can be derived by expanding the hamiltonian upto second order in $\Sigma$. Using the truncated hamiltonian in the above equation and ignoring precession due to $m_2$ and post Newtonian correction, we get the following expression: 
%
%\begin{equation}
%\Sigma = \frac{5a_1^5m_2\sqrt{a_1\mu_1}}{8a_2^3m_0J_2R_0^2} + \frac{a_1^4\mu_1}{6Gm_0J_2R_0^2}\sqrt{\frac{\mu_2}{a_2^3}}-\frac{\sqrt{\mu_1a_1}}{4}
%\end{equation}
%From the above equation we can see that the eccentricity of the particle can be increased by either adiabatically migrating the binary inward while keeping the semi-major axis of the binary constant or by increasing the semi-major axis of the binary adiabatically while keeping the binary orbit constant. 

We will now look at eccentricity corresponding to the location of stable points at $\sigma=\pi/2$. It should be noted that parameter $\eta$ is useful only at low eccentricities ($e_1 < 0.3$). Hence in the following analysis, we present our results in terms of the orbital elements of various objects in the system. The left panel of Figure \ref{fig:stbpts} shows eccentricities of stable points of the Hamiltonian as a function of semi-major axis of the companion ($a_2$)  for different values of $J_2R_d^2$. %\li{please add the corresponding $\eta$ in the plot to be more consistent with Fig. 2} 
It should be noted that the semi-major axis of the binary is kept constant in this calculation. We can see that starting from a near circular orbit, the eccentricity of the binary can be excited by migrating the binary towards $m_2$ (reducing $a_2$). But there is a limit to which the semi-major axis of the companion can be reduced without disrupting the binary.  This limit is related to the Hill radius of the binary and is given by the following expression:
\begin{equation}
a_{2,min}=2(3m_2/(m_0+m_1))^{1/3}a_1. \label{eqn:eqhill}
\end{equation}
(e.g., \cite{grishin_generalized_2017}). The minimum semi-major of the companion is shown in the figure as solid black lines. By looking at different lines we can see that the maximum eccentricity depends on the value of $J_2R_d^2$. The maximum eccentricity is higher when $J_2R_d^2$ is lower. For instance, eccentricity can be excited to $1-e \approx 10^{-3}$ for $J_2R_d^2=10^{-5}$ AU$^2$. 

\begin{figure}
	\centering
	\includegraphics[width=1.\linewidth,height=0.6\linewidth]{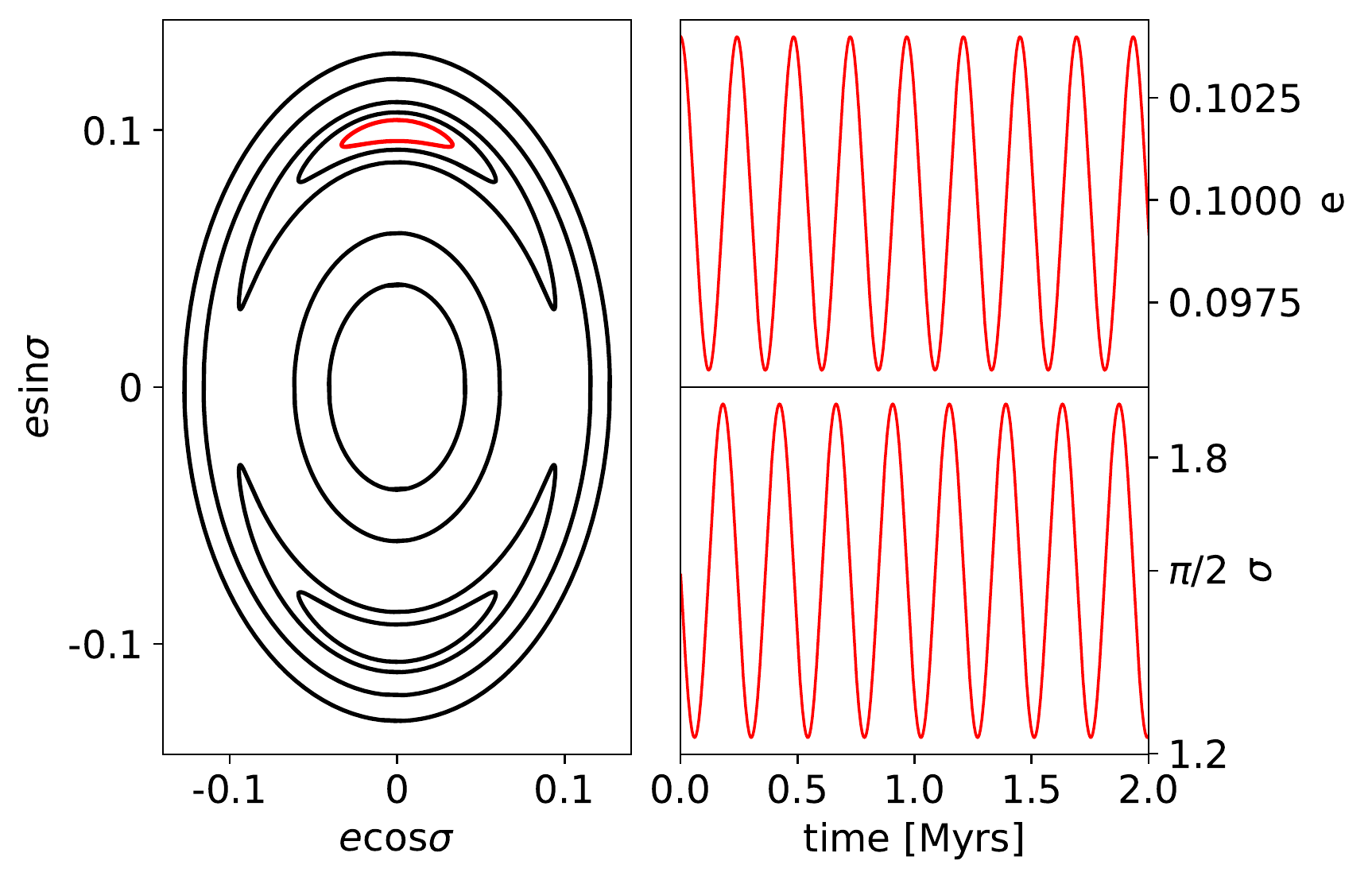}
	
	\caption{Numerical solution to the secular equations of motion  (Eqn. \ref{eqn:eqcomp}). The left panel shows the contours of the Hamiltonian from the numerical solution to Eqn. \ref{eqn:eqcomp}. Panels on the right show the evolution of eccentricity and resonant angle ($\sigma$) corresponding to the red contour shown on the left. We use the following parameters to make this plot: $a_1 = 1 \text{ AU },J_2R_d^2 = 10^{-4}\text{AU}^2(J_2=4\times10^{-4}),m_0 = 60M_\odot, m_1 = 60M_\odot, m_2 = 10^6M_\odot$ and $a_2=10911.3 \text{ AU}.$ It corresponds to $\eta=-63.8$. }
	\label{fig:numex}
\end{figure}

In the right panel of Figure \ref{fig:stbpts} we change the mass of binary components (dotted lines) and the companion (solid lines). We can see that the maximum eccentricity decreases as the mass of the binary increases. On the other hand it can be shown  that (by using Eqn. \ref{eqn:eqhill} in Eqn. \ref{eqn:sbeq}) the maximum eccentricity at the Hill radius is independent of the mass of the companion. Hence the maximum ecccentricity corresponding to the solid cyan line ($m_2 = 10^8$M$_\odot$) in the right panel is same as the dashed cyan line in the left panel ($m_2 = 10^6$M$_\odot$).
%The right panel of Figure $\ref{fig:stbpts}$ shows locations of stable points of the hamiltonian as a function of semi-major axis of the binary ($a_1$). Semi-major axis of the companion is kept constant in this calculation. We can see that the eccentricity can be excited by increasing the semi-major axis of the binary. Similar to the  left panel, the maximum eccentricity is higher when $J_2 R_0^2$ is smaller.

\section{Numerical results}
\label{sec:numres}
\subsection{Numerical Integration}
\label{sec:numint}
To further study the dynamics of a binary trapped in evection resonance, we numerically solve the equations of motion. More specifically, we use the 4th order Runge-Kutta method implemented in GNU scientific library. 
%For our simulations we found a relative error tolerance of $10^{-6}$ and an absolute error tolerance of $10^{-8}$ to be adequate. 0
A variable timestep scheme is used to make sure that the orbital elements of the binary do not change by more than 1 percent in a single timestep.

\begin{figure*}
	\centering
	\includegraphics[scale=0.8]{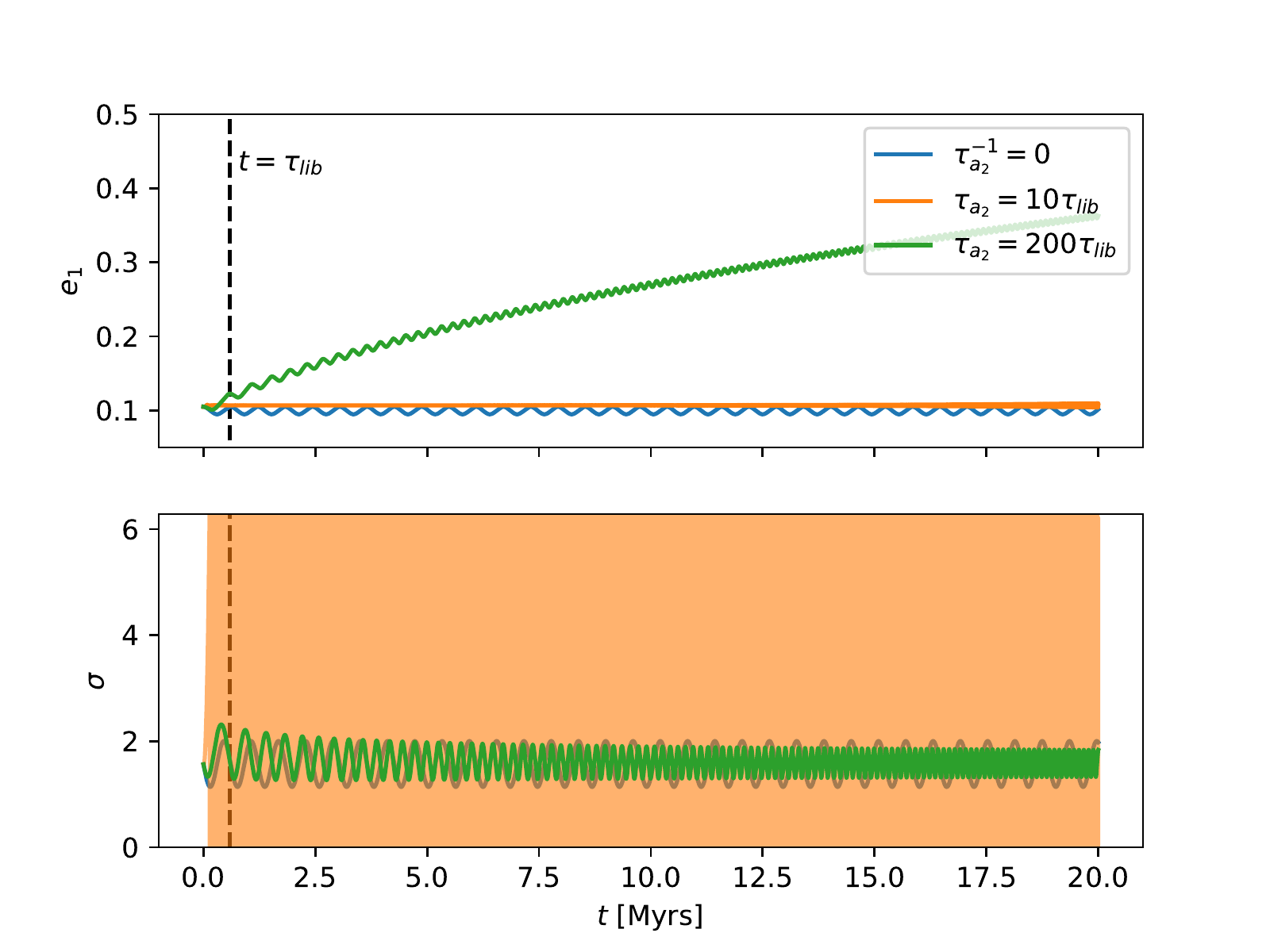}
	\caption{Time evolution of the system with three different migration timescales, $\tau_{a_2}$. We can see that in a no migration scenario, the resonant angle librates and the eccentricity of binary undergoes small oscillations around 0.1. When binary migrates fast, it's eccentricity is increased and is disrupted within million years (not shown). For $\tau_{a_2}=10\tau_{lib}$, binary leaves the resonance and the eccentricity is not excited. Finally, when the migration is slow enough ($\tau_{a_2}=200\tau_{lib}$), binary sweeps through resonance and the eccentricity is increased. We use the following parameters to make this plots: $a_1 = 1 \text{ AU },J_2R_d^2 = 10^{-4}\text{ AU}^2 (J_2=4\times10^{-4}),m_0 = 10M_\odot,m_2 = 10^6M_\odot,m_1 = 10M_\odot$ and $a_2=20336.4 \text{ AU}.$}
	\label{fig:migration}
\end{figure*}

Figure \ref{fig:numex} shows numerical solution to Eqn. \ref{eqn:eqcomp} for the initial conditions mentioned in the caption. In the panel on the left, the numerical solution is shown in red in $e\cos\sigma-e\sin\sigma$ space. Contours of the Hamiltonian are also shown for comparison. The evolution of eccentricity ($e$) and resonant angle ($\sigma$) are shown in the two panels on the right. We can see that the resonant angle librates around $\pi/2$ and  eccentricity oscillates near it's initial value of 0.1. In addition, the left panel shows that the numerical solution occupies one of the contours of the Hamiltonian which is a conserved quantity.

So far we have only looked at the dynamics of a three body system. In the following we describe additional effects like disk migration and gravitational radiation which are needed for the binary to merge. These effects are included in our ensemble simulations which are discussed in subsection \ref{sec:mrgtsc}-\ref{sec:prob}.

\subsection{Disk Migration}
\label{sec:secmig} 
In our setup we assume that the binary is embedded in an AGN disk which causes the orbit of the binary as well as the orbit around companion to decay. To model the migration we use the following prescription from \cite{lee_dynamics_2002}:
\begin{eqnarray}
\frac{\dot{a}_1}{a_1} = -\frac{1}{\tau_{a_1}}, \nonumber \\
\frac{\dot{e}_1}{e_1} = -\frac{1}{\tau_{e_1}}, \nonumber \\
\frac{\dot{a}_2}{a_2} = -\frac{1}{\tau_{a_2}}. \label{eqn:eqmig}
\end{eqnarray} 
Where $\tau_{a_1}$, $\tau_{e_1}$ and $\tau_{a_2}$ are model parameters which depend on the AGN disk. For the eccentricity to be excited, the system should remain in resonance and migrate adiabatically. Hence the migration timescale should be much longer than the libration timescale of the system which at low eccentricites is given by:
\begin{equation}
\tau_{lib} = \frac{2\pi}{\sqrt{1-\eta}}\frac{2n_1a_2^3}{15Gm_2} \label{eqn:tlib}
\end{equation} 
\citep{xu_migration_2016}. 

%{\color {green} Specify what equations are being solved and what method is being used.} 
We now add disk migration (Eqns. \ref{eqn:eqmig}) to our numerical model and solve the equations of motion using the method outlined in Section \ref{sec:numint}. %\li{please give the equation numbers.} 
Figure \ref{fig:migration} shows the evolution of a system initially trapped in evection resonance but migrating with three different values of $\tau_{a_2}$. We set $\tau^{-1}_{a_1}$ and $\tau^{-1}_{e1}$ to zero for this calculation, and we set $J_2 R_d^2$ to be constant.  In  this limit, when the binary is not migrating, we can see that $\sigma$ librates near $\pi/2$ and eccentricity undergoes small oscillations around its initial value of $0.1$. When the binary is migrating faster than the libration timescale, it quickly leaves the resonance and is disrupted by the companion. In the intermediate regime when $\tau_{a_2}=(10-100)\tau_{lib}$, binary escapes resonance without any eccentricity excitation. Finally, when the binary is migrating on a timescale of $200 \tau_{lib}$, we can see that the binary stays in resonance and the eccentricity is excited. Dashed lines show the libration timescale from Eqn. \ref{eqn:tlib}. Hence the migration time needed to excite the eccentricity depends on the libration timescale of the system.

\begin{figure}
	\centering
	\includegraphics[width=1.0\linewidth,height=0.50\linewidth]{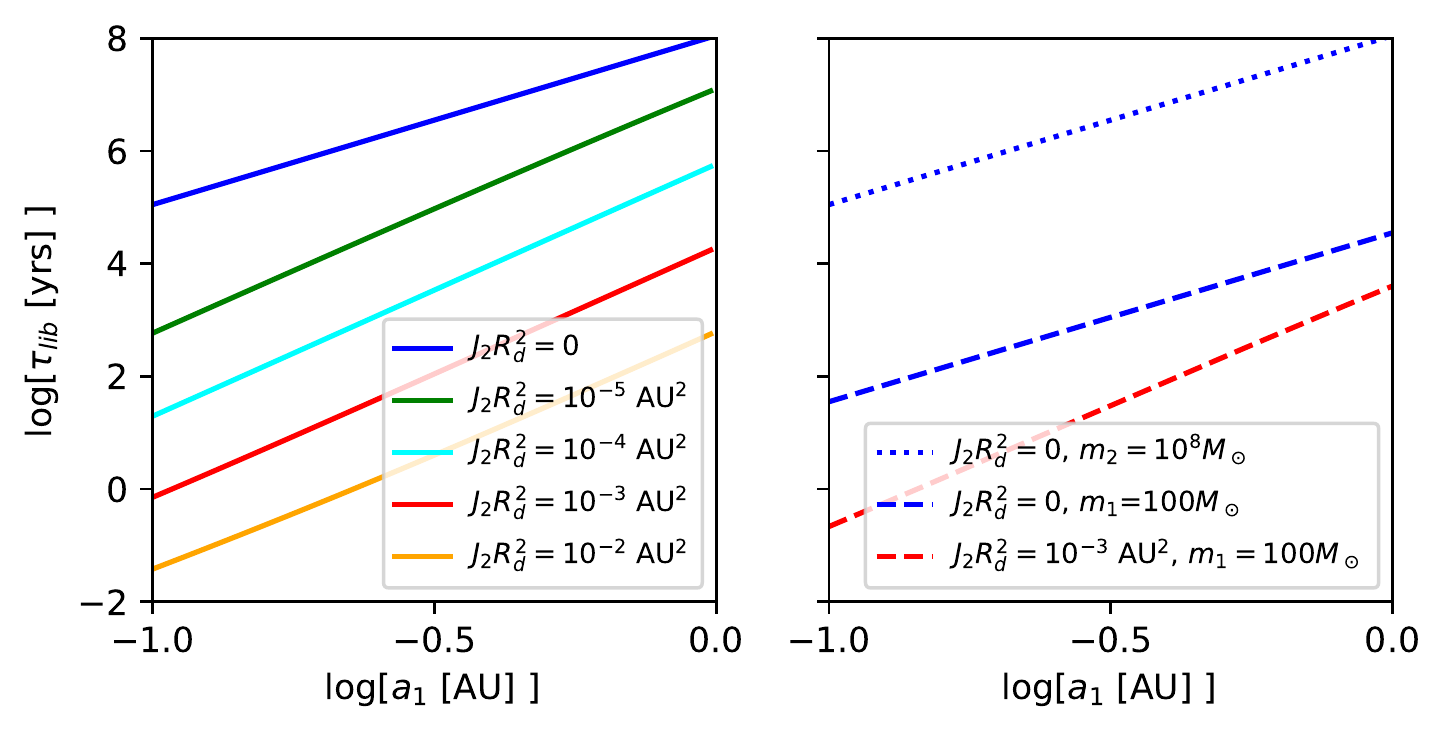}
	\caption{The left panel shows the libration timescale as a function of semi-major axis of the binary. We can see that the libration timescale increases with the semi-major axis of the binary. Also, by looking at different lines, we find that the libration timescale decreases as $J_2R_d^2$ increases. We use the following parameters to make these plots: $m_{0,1} = 10M_\odot$ and $m_2 = 10^6M_\odot$. We choose the semi-major axis of the companion ($a_2$) such that the system is captured in evection resonance at $e_1=0.01$. In the right panel we change the mass of the companion (dotted) and the mass of the binary (dashed). Other parameters are same as the left panel i.e. for the dotted line, is $m_0=m_1=10 M_\odot$ and for the dashed lines $m_2 = 10^6 M_\odot$ and $m_0=m_1$. We can see that the libration timescale decreases as the mass of the binary mass increases. In addition, libration timescale does not depend on the mass of the companion.}
	\label{fig:libts}
\end{figure}

\begin{figure*}
	\centering
	\includegraphics[width=1.0\linewidth,height=0.4\linewidth]{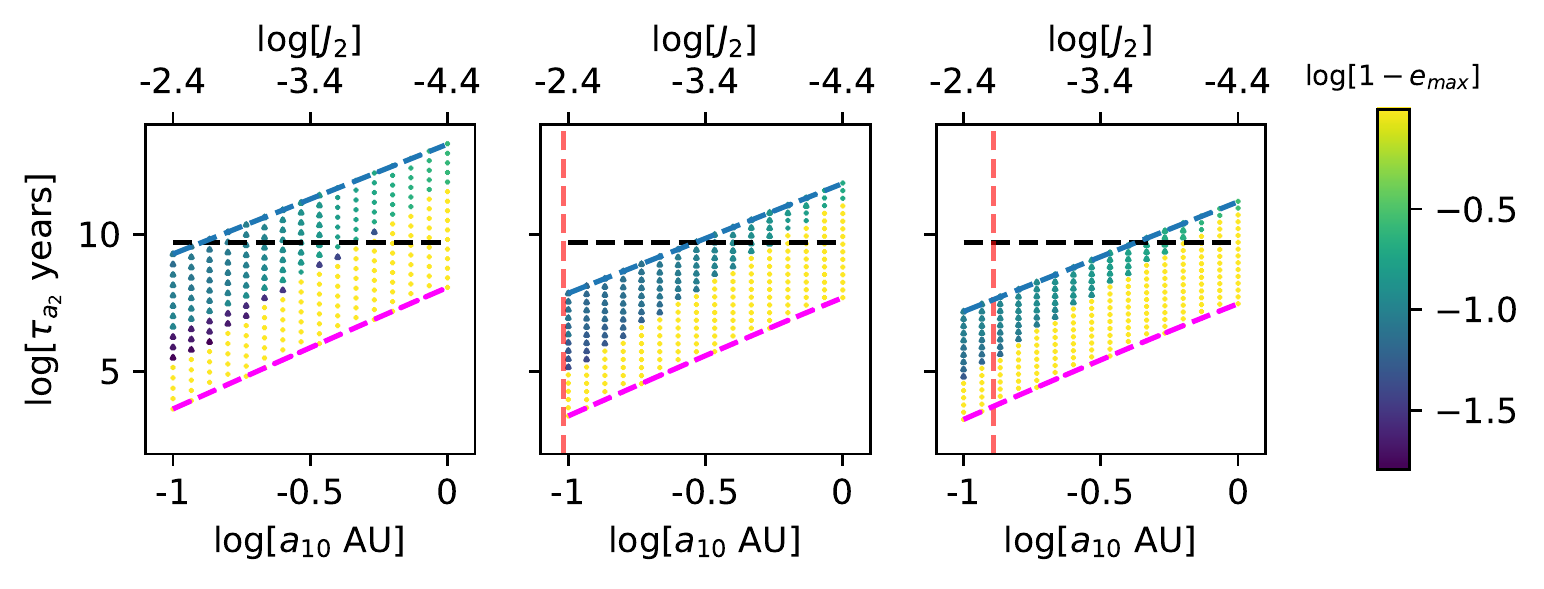}
	
	\caption{Results of ensemble simulations: distribution of maximum eccentricity for equal mass binaries. The binary is migrating on a timescale shown on the y-axis and the initial semi-major axis of the binary is shown on the x axis.  Color shows the maximum eccentricity obtained in the simulation. The three panels correspond different masses of binaries with $m_0=m_1=$: $20 M_\odot$ (left), $60 M_\odot$ (middle) and $100 M_\odot$ (right). Simulation end time of 5 billion years is shown as black dashed line. Migration time is sampled between libration time (magenta line) and GR merger time (blue). Regions where binaries merge are shown as triangles and where they don't are shown as circles. We can see that when the migration time is 10-100 times the libration time, eccentricity is excited and binaries merge. Vertical dashed red lines show the sum of radii of main sequence  binary stars with masses of 60+60$M_\odot$ (middle panel) and 100+100$M_\odot$ (right panel) respectively. 
	%In addition, solid red lines mark the region where main sequence binary stars can collide with each other i.e. $(1-e_{max})a_{1}<2R_*$. 
	We use a variable $J_2$ prescription as described in Eqn. \ref{eqn:varj2}.  We use the following parameters to get these results: $J_2(0)R_d(0)^2=10^{-5} \text{AU}^2 ,m_2=10^6M_\odot$ 
	%{\color {green} what is the initial $a_2$?} 
	and initial binary eccentricity $e_{10}=0.01$.  The value of initial $a_2$ is chosen to make sure the binary is trapped in resonance: for 20+20$M_\odot \in \{351.0,69191.8\}$AU, 60+60$M_\odot \in \{238.5 ,40843.41\}$AU and 100+100$M_\odot \in \{197.2065
,30263.47\}$AU.
	%{\color {red} It may be  useful to draw a line to indicate where two main sequence binary stars may touch each other at perigee, with $(1-e_{\rm max}) a_1 \lesssim R_\ast(m_0) + R_\ast (m_1)$.}  
	}
	\label{fig:en_a2a}
\end{figure*}

We will now look at the dependence of $\tau_{lib}$ on the semi-major axis of the binary $a_1$. Using Eqn. \ref{eqn:tlib}, the left panel of Fig. \ref{fig:libts} shows $\tau_{lib}$ as a function of $a_1$ for different values of $J_2R_d^2$. We can see that the libration timescale decreases with $J_2R_d^2$. This means that the migration rate needed to keep the binary in resonance also increases as $J_2R_d^2$ decreases. Hence, despite the fact that smaller values of $J_2R_d^2$ allow for greater eccentricity excitation (See Fig. \ref{fig:stbpts}), they may not be ideal due to the time needed to excite the eccentricity. For instance, in a no disk scenario ($J_2=0$) with $a_1=1$ AU, the libration timescale is more than 100 million years. When GR precession dominates, it can be shown that, $\tau_{lib} \propto a_1^3$ and also when $J_2$ precession dominates $\tau_{lib} \propto a_1^{9/2}$. In the right panel we change the mass of the companion (dotted lines) and the mass of the components of the binary (dashed lines). We can see that the libration time decreases with the mass of the binary. In addition, the libration time does not depend on the mass of the companion.

%\li{I think it is smoother to move the following paragraphs here:}

 How does the migration timescales needed for evection resonances compare with typical AGN disk migration timescales? Depending on the mass of the binary, it can migrate by either Type I or Type II migration mechanism. Low-mass objects embedded in a gaseous disk undergo Type I migration, which is caused by Lindblad and co-rotation torques exerted by the disk on the object (e.g., \cite{goldreich_excitation_1979}).  Additionally, more massive objects open a gap in the disk and undergo usually slower Type II migration (e.g., \cite{lin_tidal_1993}). To open a gap in the disk, the torques from the object must overcome the viscous torques in the disk. In geometrically-thin AGN disks, \cite{baruteau_binaries_2010} find that objects with $m \gtrsim 10 M_\odot$ may undergo Type II migration. 
Both Type I and Type II migration timescales and directions depend on the disk's surface density and entropy 
distribution \citep{paardekooper2010, paardekooper2011, chen2020}. 
Multiple steady-state AGN disk models have been explored in literature. For instance, \cite{sirko_spectral_2003} assume a thin $\alpha$ disk with a constant accretion rate. A feedback mechanism  like star formation is invoked to prevent the AGN disk from fragmenting by maintaining a Toomre parameter ($Q$) of order unity. In an alternative model explored by \cite{thompson_radiation_2005}, angular momentum transport is assumed to take place via bar or spiral waves on a global
scale. Using the standard $\alpha$-prescription for local turbulent viscosity in AGN disks, the migration timescales for single stars/BHs are 
estimated to be, $\tau_{\text{Type I}} \sim 10^7$ years and $\tau_{\text{Type II}} \sim 10^8$ years \citep{baruteau_binaries_2010,levin_starbursts_2007}. 
These timescales may be significantly modified by the vigorous gravito-turbulence \citep{rowther2020}. 
Moreover, global structure such as bars, spiral pattern, 
or warps may either promote or thwart  migration and potentially lead to a concentration of stars and BHs near migration traps. 
However, the coexistence of multiple nearby stars/BHs introduces 
mutual perturbation, disturbs horseshoe gas streamlines around their orbits, 
and interferes their torque on the disk gas through Lindblad and corotation resonances.
For the purpose of this study, we assume that the migration rate of a binary is of same order of magnitude as the migration rate of a single object with the combined mass. Hence, the proposed mechanism  works in the parameter space where the $\tau_{lib} \lesssim 10^6$ years. From Figure \ref{fig:libts}, we can see that low mass binaries ($\sim 10 M\odot$) with massive disks ($J_2R_d^2 > 10^{-4}$ AU$^2$) or high mass binaries ($\sim 100 M_\odot$) could be captured in evection resonance and have their eccentricity excited.

If the gas disk is isothermal, the migration as assumed in this work is always inwards.  However, in some adiabatic disk models, 
the migration can be outwards \citep{paardekooper2010, paardekooper2011, chen2020}.  Although it is possible for AGN disks to 
have migration traps where the net torques on objects embedded in the disk changes sign \citep{bellovary_migration_2016,
secunda_orbital_2020}, the robustness and location of such traps depends on the disk model.  If an isolated binary is captured in such 
a migration trap, it may not be able to enter the evection resonance. However, dynamical perturbation by other nearby trapped 
stars/BHs may excite its eccentricity as well as modify its resonance torque on the disk gas, migration direction, 
and timescale.  In view of these uncertainties, 
we rather use a simple model in this paper, for binary migration and binary hardening in AGN disks. To account for the range of 
possibilities in the AGN disk models we run simulations for a wide range of migration/hardening timescales.

%\subsection{Gravitational radiation}
%\label{sec:secgrad}

%This can be seen in Figure, where Eqns. \ref{eqn:grrad} are solved numerically to show the merger timescale as a function of semi-major axis of the binary. Different lines correspond to different initial eccentricities. 
%\li{This is not smooth to have a section on GW here, let us merge it to the next section}

\subsection{Eccentricity Excitation}
\label{sec:mrgtsc}
We next discuss the efficiency of evection resonances in eccentricity excitation of the binaries after including gravitational radiation. We allow $J_2 R_d^2$ to vary assuming a constant disk density profile in the simulations for a more realistic disk.
In addition to interactions with the disk, gravitational radiation can also cause the binary orbit to shrink. This effect 
is especially important when the eccentricity is excited. The secular evolution of semi-major axis and eccentricity of the binary is given by \cite{peters_gravitational_1964}:
\begin{eqnarray}
\frac{da_1}{dt} &=& -\frac{64}{5}\frac{G^3m_0m_1(m_0+m_1)}{c^5a_1^3(1-e_1^2)^{7/2}}(1+\frac{73}{24}e_1^2+\frac{37}{96}e_1^4), \nonumber \\ 
\frac{de_1}{dt} &=& -\frac{304}{15}e_1\frac{G^3m_0m_1(m_0+m_1)}{c^5a_1^4(1-e_1^2)^{5/2}}(1+\frac{121}{304}e_1^2).  \label{eqn:grrad}
\end{eqnarray}
The decay timescale associated with this mechanism for a circular orbit is given by:
\begin{equation}
t_{GW} = \frac{a_1^4}{4}\frac{5}{64} \frac{c^5}{G^3m_0m_1(m_0+m_1)} \label{eqn:tmerg}.
\end{equation} 
When the binary is eccentric, the decay timescale for an eccentric binary, with the same $a_1$, is further reduced by a factor of $(1-e_1^2)^{7/2}$. 

%{\color {green} Specify what equation are being solved and what method is used here.}
Next, we numerically solve the equations of motion using the method outlined in section \ref{sec:numint} and include perturbations from the companion (Eqn. \ref{eqn:eqcomp}), migration due to AGN disk (Eqn. \ref{eqn:eqmig}) and gravitational radiation (Eqn. \ref{eqn:grrad}). 
%We run two sets of simulations. In our first set of simulations we assume that the orbit of the binary does not decay due to the AGN disk i.e. $\tau^{-1}_{a1},\tau^{-1}_{e1}=0$. In our second set of simulations, the AGN disk is allowed to damp and circularize the binary orbit. 
Initial conditions are chosen to ensure that the binary starts near evection resonances. This is done by solving Eqn. \ref{eqn:sbeq} for the semi-major axis of the companion ($a_2$). Whenever the system leaves the resonance (i.e., the resonant angle stops librating), we continue our simulation using only Eqns. \ref{eqn:eqmig} and \ref{eqn:grrad}. This is done to reduce computational time. All the simulations are run for a maximum time of 5 billion years and all binaries whose semi-major axis shrinks to $10^{-4}$ AU in 5 billion years are assumed to have merged.

Figure \ref{fig:en_a2a} show results of our secular simulations as the binary migrates towards the supermassive BH in the AGN disk. Initial semi-major axis of the binary is shown on the x axis and the migration timescale ($\tau_{a_2}$%{\color {green} do you mean $\tau_{a_2}$?}
) is shown on the y axis. We neglect the hardening of the binary components or their eccentricity damping in this set of simulations (i.e., $\tau_{a_1}^{-1}$ and  $\tau_{e_1}^{-1}$ are taken zero ). Simulation end time of 5 billion years is shown as black dashed line. 
%({\color{green} It would be nice if we can have an approximate analytic formula for $1-e_{\rm max}$ as a function of $a_1, m, \tau_{a_2}$}).

For a given initial semi-major axis, the migration time is sampled between the libration time (Eqn. \ref{eqn:tlib}) and the merger time (Eqn. \ref{eqn:tmerg}) shown as magenta and blue dashed lines  respectively. Regions marked using triangles show the parameter space where binaries merge and those marked using circles show regions where they don't. Color shows the maximum eccentricity attained before the end of the simulation. We can see that the region above the black dashed line marks parameter space where migration rate is too slow for mergers to happen.  
%It should be noted that merger time is proportional to the migration time shown on the y-axis. 
We can also see that the fastest merger time is around 10-100 times the libration time of the system. If the binary is migrating faster, there is no eccentricity excitation and hence the binary doesn't undergo a merger. 

Due to evection resonances, eccentricity can be excited to $1-e_{\rm max} > 10^{-1.5}$ before gravitation radiation becomes dominant and circularizes the orbit. 
Comparing different panels we can see that it is easier for heavier binaries to merge. For instance, while a $20+20 M_\odot$ binary 
%({\color {green} is this mass  =  $m_0=m_1=20 M_\odot$ ?})
starting with a semi-major axis of 0.75 AU cannot merge, both $60+60 M_\odot$ and $100+100 M_\odot$  binaries starting with the same semi-major axis can merge in 5 billion years. Also, it should be noted that this can happen only if their eccentricity is excited. 

%{\color {green}  
These  results apply to both binary black holes and  binary stars with radii $R_0$ and $R_1$. The latter systems would merge when $a_1 (1-e_{\rm max}) \leq (R_0 + R_1)$, and thus it requires a lower maximum eccentricity for the mergers to occur.
%} 
%Thus, we also plot the region of parameter space where main-sequence binary stars of equivalent mass can collide due to eccentricity excitation. 
In Figure \ref{fig:en_a2a}, vertical dashed red lines show the sum of radii of main sequence binary stars with equivalent masses \footnote{We assume a mass-radius relationship of $R_*/R_\odot = (M/M_\odot)^{0.57}$}.  In our variable $J_2$ simulations we find that in the parameter space where blackhole-binaries merge, equivalent main-sequence stars would always collide with each other. 
%and solid red lines mark the region where main sequence binary stars can collide with each other i.e. $(1-e_{max})a_{1}<2R_*$. All binary main-sequence stars below the solid red line would collide and merge. We can see that merger of main sequence binary stars is allowed in most of the parameter space except when the the binaries migrate slowly($\tau_{a_2} \sim 10^9$ yrs). 

\begin{figure}
	\centering
	\includegraphics[width=1.0\linewidth,height=0.75\linewidth]{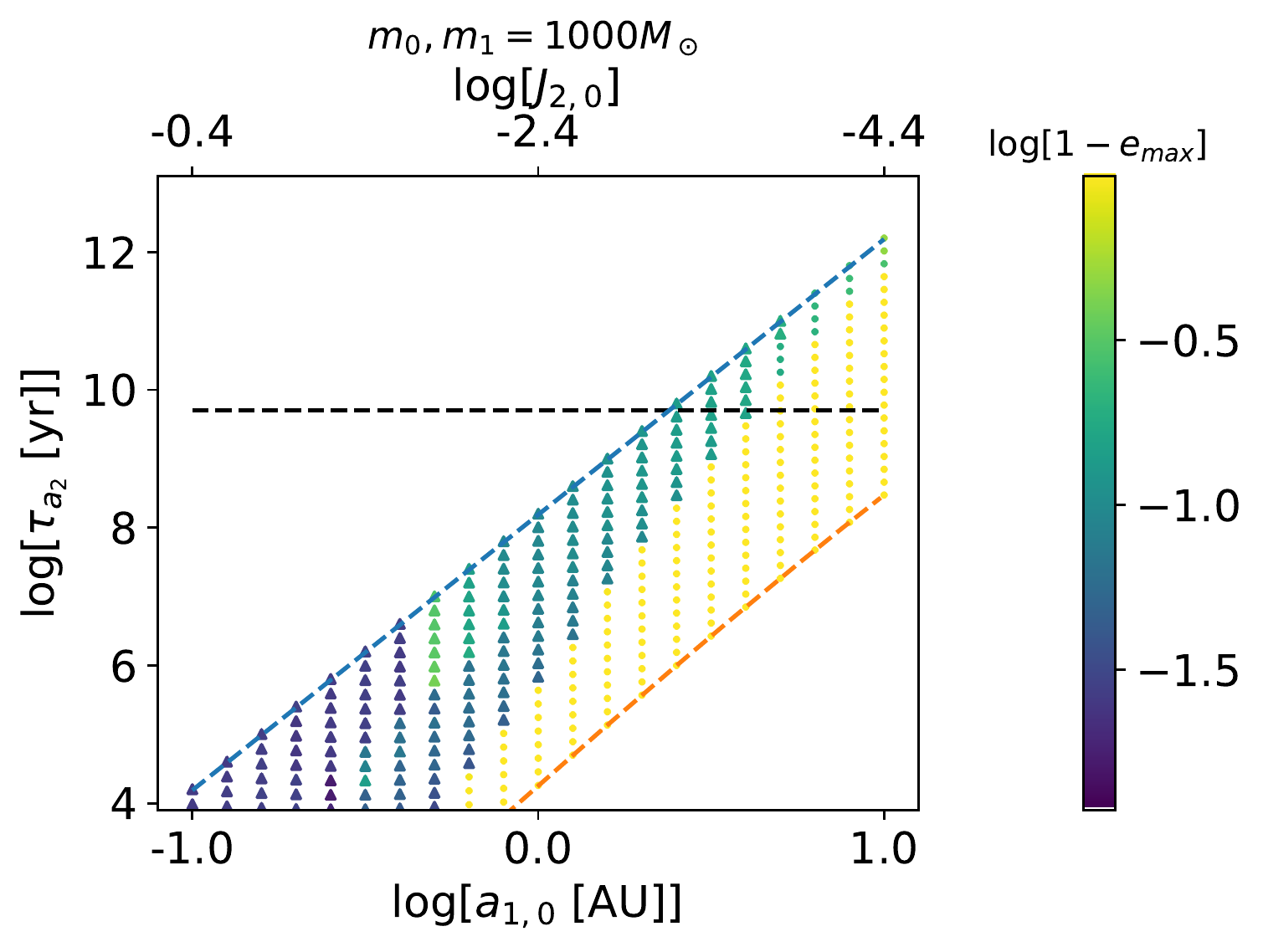}
	
	\caption{Results of ensemble simulations for massive stars in AGN disks: distribution of maximum eccentricity for equal mass binaries. The binary is migrating on a timescale shown on the y-axis and the initial semi-major axis of the binary is shown on the x axis.  Color shows the maximum eccentricity obtained in the simulation.  Simulation end time of 5 billion years is shown as black dashed line. Migration time is sampled between libration time (orange line) and GR merger time (blue). Regions where binaries merge are shown as triangles and where they don't are shown as circles. In this regime, we can see that binaries remain in resonance for most of the parameter space and the eccentricity of the binary is excited. 
	%But for $\tau_{a_2} \lesssim 10^5$ yrs, binaries are disrupted before can they merger due to gravitational radiation. 
	We use the following parameters to get these results: $m_2=10^8M_\odot$ %{\color {green} what is the initial $a_2$?} 
	and initial binary eccentricity $e_{10}=0.01$. We use a variable $J_2$ prescription with $p=0$ as described in Eqn. \ref{eqn:varj2}. The initial $J_2R_0^2=10^{-3}$ AU$^2$  The value of initial $a_2$ is chosen to make sure the binary is trapped in resonance: $a_2 \in \{23.1,651967.8\}$ AU }
	\label{fig:en_a2}
\end{figure}

%\li{
AGN stars can rapidly become very massive ($M > 100$M$_\odot$) in AGN disks \citep{Cantiello21}, and the higher mass of stars allows shorter libration timescale, and thus it's easier to capture the stars in evection resonances. For illustration, we included results for a stellar binary with equal mass stars of $1000$M$_\odot$ in Figure \ref{fig:en_a2}. Figure \ref{fig:en_a2} shows the maximum eccentricities reached within $5$ Gyr with different binary semi-major axes and binary ($a_2$) migration timescales. 
%The libration timescales are all lower than the $a_2$ migration timescales, and thus the stars could be captured in the resonance. 
The triangles represent successful mergers within 5 Gyrs. Evection resonances are efficient in producing mergers, where most of the runs lead to mergers with migration timescales above $\gtrsim 10^4$ yrs. 
%}

%\li{Hareesh: could you please cut the y-axis above $10^4$yrs?}

%{\color {green} 

\begin{figure}
	\centering
	\includegraphics[width=1.0\linewidth,height=0.55\linewidth]{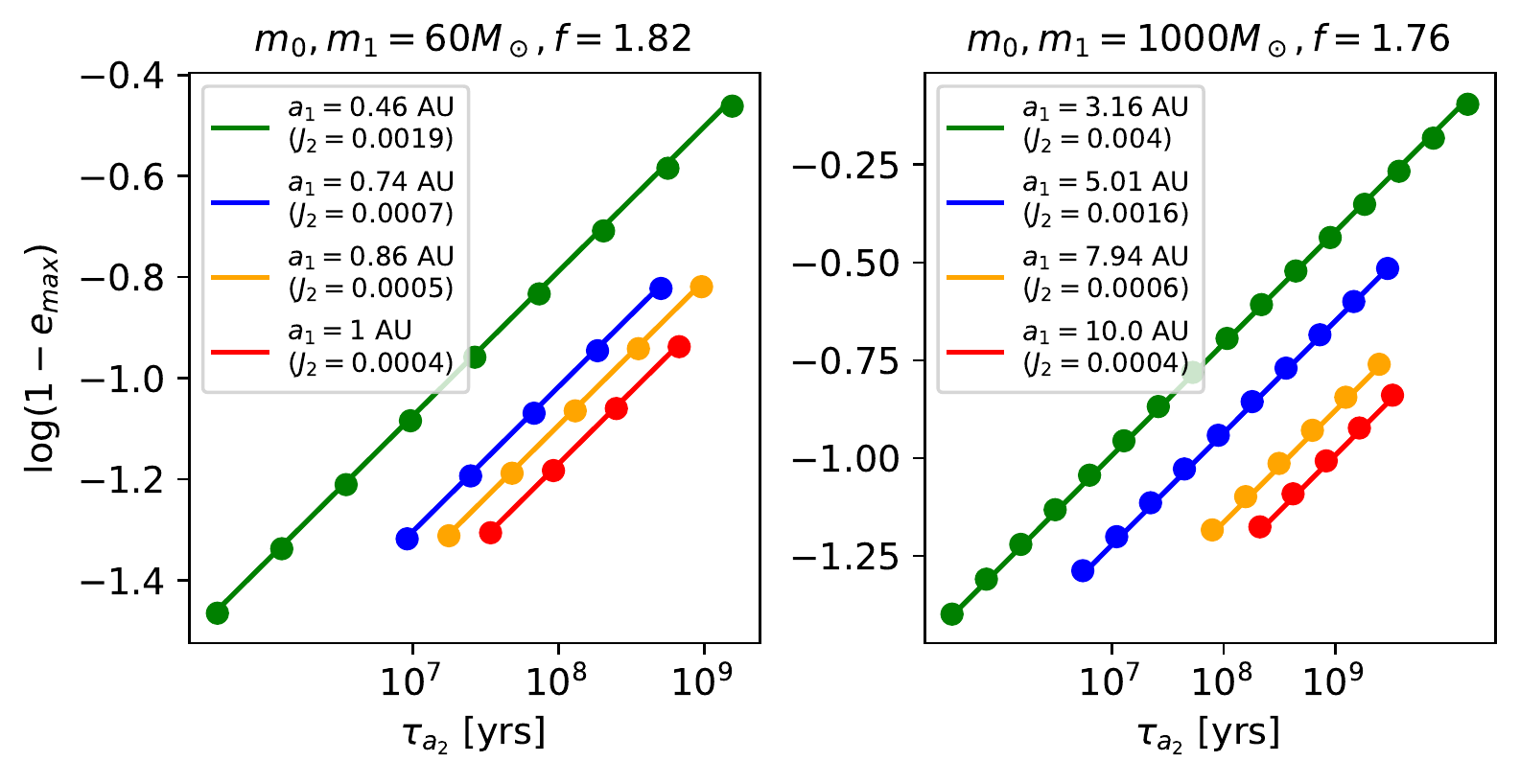}
		\caption{Comparison of analytical estimate of $e_{max}$  (Eqn. \ref{eqn:emaxan}) with direct secular simulations. The orbital decay timescale ($\tau_{a_1}$) of the binary is shown on the x axis and the maximum eccentricity ($1 - e_{max}$) is shown on the y axis. Results from simulations are shown using as filled circles. Different colors correspond to different initial semi-major axes as shown in the legend. The left panel shows results for 60+60 $M_\odot$ binaries and 1000+1000 $M_\odot$ binaries. The analytical estimate is shown (Eqn. \ref{eqn:emaxan}) using solid lines. Factor ($f$) of 1.82 gives the best fit for the left panel and 1.76 for the right panel. We use the following parameters to get these results: $J_2R_d^2=10^{-4} \text{AU}^2 ,m_2=10^6M_\odot$ for the left panel and $J_2R_d^2=10^{-2} \text{AU}^2 ,m_2=10^8M_\odot$ and initial binary eccentricity $e_{10}=0.1$ for the right panel.}
	\label{fig:emaxan}
\end{figure}

\begin{figure}
	\centering
	\includegraphics[width=0.9\linewidth,height=0.75\linewidth]{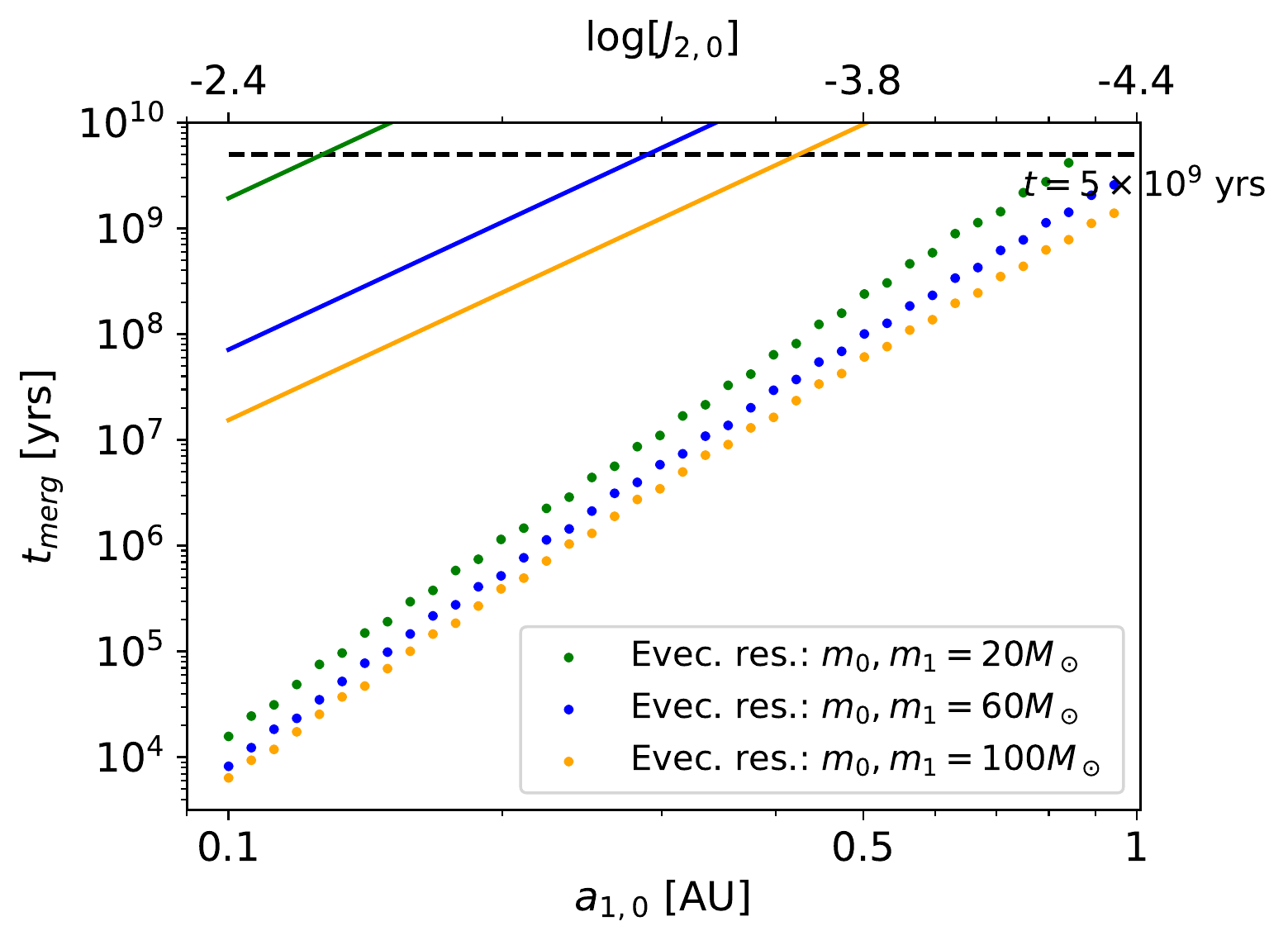}
	
	\caption{Merger time as a function of initial semi-major axis of three equal mass binaries. Filled circles show the shortest merger time for a migrating binary trapped in evection resonances. For comparision, merger time due to gravitational radiation (Eqn. \ref{eqn:grrad}) without eccentricity excitation is shown as solid lines. Due to eccentricity excitation, merger time is 3-5 orders of magnitude smaller for migrating binaries. We use the following parameters to make these plots: $J_2(0)R_d(0)^2 = 10^{-5} \text{AU}^2, m_2 = 10^6M_\odot, e_{1,0}=0.05$ and $a_2$  is chosen so that the binary starts near evection resonance. We use a variable $J_2$ prescription with $p=0$.}
	\label{fig:mtcomp}
\end{figure}

We can derive a simple analytical expression for the maximum eccentricity that a binary can achieve during an evection resonance mediated merger by comparing the rate of change of eccentricity due to evection resonance and the rate of change of eccentricity due to gravitational radiation. As mergers are expected to happen at high eccentricities, we can simplify  the condition for evection resonances (Eqn. \ref{eqn:sbeq}) in the limit $e \rightarrow 1$ as following:
\begin{equation}
    1-e \simeq \frac{3G(m_0+m_1)\sqrt{m_0+m_1}}{2a_1^{5/2}c^2\sqrt{m_0+m_1+m_2}}a_2^{3/2} \label{eqn:omej2}
\end{equation}
in a no disk scenario. When the precession due to disk is much faster than GR precession, the maximum eccentricity is given by: 
\begin{equation}
1-e \simeq \sqrt{\frac{3}{8}}\frac{\sqrt{J_2 R^2_d m_0}}{(a_1^7(m_0+m_1)(m_0+m_1+m_2))^{1/4}}a_2^{3/4} \label{eqn:omegr}.
\end{equation}
The eccentricity of a binary trapped in evection resonance and migrating towards SMBH adiabatically would be given by above expressions. Taking the derivative of above equations gives us the rate of increase of eccentricity as the binary migrates:
\begin{eqnarray}
\dot{e} &\simeq& \frac{3}{2}\frac{(1-e)}{\tau_{a_2}}  \text{  (no disk),}\nonumber \\
\dot{e} &\simeq& \frac{3}{4}\frac{(1-e)}{\tau_{a_2}}  \text{  (disk dominates).}\nonumber
\end{eqnarray}
Note that we assume that the binary separation($a_1$) and the quadrupole moment of the disk ($J_2R_0^2$ ) are constant while deriving this expression. We can now compare the above expression with Eqn. \ref{eqn:grrad} to get:
\begin{equation}
    (1-e_{max})^{7/2} = f \frac{950}{45} \frac{G^3m_0m_1(m_0+m_1)}{2^{5/2}c^5a_{10}^4}\tau_{a_2}
    \label{eqn:emaxan}
\end{equation}
where $f$ is a factor of order unity which depends on the mass of the binary. Figure \ref{fig:emaxan} compares the above equation with secular simulations. We can see that the maximum eccentricity decreases with $\tau_{a_2}$. The above expression agrees with $e_{max}$ from simulations for $f$ equal to 1.82 and 1.76 for 60+60 $M_\odot$ binaries and 1000+1000 $M_\odot$ binaries respectively.  

For main sequence stars, $R_0 \simeq R_\odot (M_0/M_\odot)^{0.6}$. Thus, for a given value $\tau_{\rm a_2}$, we can determine the critical stellar mass for the binary stars to merge.
%}{\color {green} 
This stellar application is different from the binary black hole case we have discussed earlier in this subsection. With much smaller gravitational radii, 
the relevant condition for binary black holes is determined  by the merger timescale due to gravitational radiation (Eq. \ref{eqn:grrad}).
%} 
Using Eqn. \ref{eqn:emaxan}, we get the critical mass as:
\begin{equation}
    M_{crit}=0.59M_\odot\left(\frac{a_{10}}{R_\odot}\left(\frac{f\tau_{a_2}G^3m_0m_1(m_0+m_1)}{a_{10}^4c^5}\right)^{2/7}\right)^{1.66}
\end{equation}

We note that we only consider the $J_2$ precession here. Different from the black holes, tidal bulges of the stars could also contribute to orbital precession, in addition to the effects of GR and $J_2$ due to stellar rotational flattening as well as disks \citep{liu_suppression_2015}. In particular, for synchronized rotation, orbital precession due to the tidal bulge can be much larger than that due to rotational flattening.

\subsection{Merger Timescales}
\label{sec:mrgtsc2}
We now look at the merger time of binaries. Figure \ref{fig:mtcomp} shows the merger timescale as a function of initial semi-major axis of the binary. Filled circles show the fastest merger time for migrating binaries trapped in evection resonances. For comparison, solid lines show merger time for binaries not trapped in evection resonances and undergoing orbital decay only due to gravitational radiation. Simulation end time of 5 billion years is shown as dashed black lines.  We can see that the merger time can be 3-5 orders of magnitude smaller with evection resonance. For instance, a 20+20 $M_\odot$ binary with a initial semi-major axis of 1 AU can merge within 5 billion years. The merger timescale is shorter for more massive binaries. For instance, binaries with a mass of 100 $M_\odot$ and initial semi-major axis of 0.1 AU can merge in 1000 years.  
%The dependence of merger time on the semi-major axis is roughly given by $t_{merg} \propto a^{4.5}_{1,0}$.

\iffalse
{\color {green} It would  be useful to use Eqn (\ref{eqn:tmerg}) to provide an analytic approximation for the critical eccentricity $e_{\rm merge}$  
as a function of $a_1, m,$ and $\tau_{\rm merge}$.  In order for merger to take place within some life time $\tau_{\rm life}$, 
it is necessary for $e_{\rm max} \gtrsim e_{\rm merge}$ and $\tau_{a_2} \leq \tau_{\rm merge} \leq  \tau_{\rm life}$. 
From these two conditions, we can estimate with an analytic approximation the critical value of $a_{1, \rm merger}$ as a function of 
$m$ and $\tau_{\rm life}$. (All binaries with mass $m$ caught in evection resonances with $a_1 \lesssim a_{1, merger}$ would merge 
within $\tau_{\rm life}$. From this value of  $a_{1, \rm merger}$, we can derive $a_{2, \rm merger}$, i.e. all binary with mass $m$
caught in evection resonances with $a_2 \lesssim a_{2, \rm merger}$ would merge.}
\fi

We can derive an analytical expression for the critical semi-major axis ($a_{crit}$), beyond which binaries trapped in evection resonance cannot merge within some timescale $\tau_{life}$. From Fig. \ref{fig:en_a2a}, we can see that binaries stay in resonance and have their eccentricity excited when the migration timescale is 10-100 times the libration timescale. Binaries with large semi-major axis can have libration timescale ($\tau_{lib}$) greater than $\tau_{life}$. This requirement gives us the following condition for critical semi-major axis:
\begin{equation}
    \tau_{life}=f_{lib}\tau_{lib}(a_{1,crit})
\end{equation}
Where $f_{lib}$ %{\color{green}  $f_{lib}$} 
is a factor of order 10-100. Using Eqn. \ref{eqn:sbeq}, we can solve for the semi-major axis of the binary around SMBH ($a_2$) when it is captured in evection resonance and put it in Eqn. \ref{eqn:tlib} to get the critical semi-major axis from the above equation. For low eccentricity capture this gives us:
\begin{eqnarray}
a_{1,crit}&=& \left( \frac{9\sqrt{5}e_0}{2\pi f_{lib}}\right)^{1/3} (c\tau_{life}R^2_{G,1})^{1/3}\nonumber \\
&=& 1.21 \left[\left( \frac{e_0}{0.1} \right) \left( \frac{\tau_{life}}{10^{10} \text{ yrs}} \right)
\left( \frac{m_0+m_1}{30 M_\odot} \right)^{2} \left( \frac{10}{f_{lib}} \right)\right]^{1/3} \text{ AU}\nonumber
\end{eqnarray}
when there is no disk and when disk dominates critical semi-major axis is given by:
\begin{eqnarray}
a_{1,crit}&=& \left(\frac{405 e_0^{2/9}}{16 (f_{lib}\pi)^{2/9}}\right) \left(q^3 R^7_{G,1}(c\tau_{life})^2 \alpha_G \right)^{1/9} \nonumber \\
&=& 5.38 \left(\left(\frac{J_2R_d^2}{10^{-4}\text{ AU}^2}\right) \left(\frac{q_1}{0.5}\right  )       \left(\frac{e_0}{0.1}\right)^2 \right)^{1/3} \nonumber \\
&& \times \left( \left(\frac{\tau_{life}}{10^{10}\text{ yrs}}\right)^2 \left(\frac{m_0+m_1}{30 M_\odot}\right) \left(\frac{10}{f_{lib}}\right)^{2}\right)^{1/3} \text{AU}\nonumber
\end{eqnarray}
Here, $e_0$ is the initial eccentricity, $r_{G,1}=G(m_0+m_1)/c^2$ is the gravitational radius of the binary, $r_{G,2}=Gm_2/c^2$ is the gravitational radius of the $m_2$, $\alpha_G=J_2R_d^2/r^2_{G,1}$ is the disk mass ratio and $q_1=m_0/(m_0+m_1)$ is the mass ratio of the binary. The above expressions can also be derived using $\dot{\varpi}(a_{1,crit})\tau_{life} \sim 1$, which tells us the semi-major axis where precession is too slow. It should be noted that binaries with semi-major axis less than $a_{1,crit}$ can merge if they are migrating slowly  ($\tau_{a_2}>f_{lib}\tau_{lib}$). On the other hand, binaries with semi-major axis greater than $a_{1,crit}$ cannot merge within $\tau_{life}$. From the above expressions we can get the corresponding distance from the SMBH:
\begin{eqnarray}
a_{2,crit}&=& \left(\frac{2025\sqrt{5}}{(2\pi)^5}\right)^{1/9}\left(\frac{e_0}{f_{lib}}\right)^{5/9}\left(R_{G,1}R^3_{G,2}(c\tau_{life})^5\right)^{1/9} \nonumber\\
&=& 2.32\left[\left(\frac{m_0+m_1}{30 M_\odot}\right)^{1/3} \left(\frac{e_0}{0.1}\right)^{5} \frac{m_2}{10^6 M_\odot}\right]^{5} \nonumber \\  
& &\times \left(\frac{\tau_{life}}{10^{10} \text{ yrs}}\right)^{5/3}  \left(\frac{10}{f_{lib}}\right)^{5/3} \text{ pc} \nonumber 
\end{eqnarray}
when there is no disk and
\begin{eqnarray}
a_{2,crit}
&=& \left(\frac{3^{10}5^7e_0^{14}}{2^{10}\pi^{14}f_{lib}^{14}}\right)^{1/27} \left(\alpha_G q_1q_2^3R^{13/3}_{G,1}(c\tau_{life})^{14/3}\right)^{1/9} \nonumber \\
&=& 4.81 \left(\frac{J_2R_d^2}{10^{-4} \text{AU}^2} .\frac{m_0}{30M_\odot}\right)^{1/9} \left(\frac{m_2}{10^6 M_\odot}\right)^{1/3} \nonumber \\
&&\times \left(\left(\frac{e_0}{0.1} \frac{\tau_{life}}{10^{10} \text{ yrs}}\frac{10}{f_{lib}}\right)^{14} \left(\frac{30 M_\odot}{m_0 + m_1}\right)^5\right)^{1/27} \text{pc} \nonumber
\end{eqnarray}
when precession due to disk dominates. Here $q_2=m_0/m_2$.

\begin{figure}
	\centering
	\includegraphics[width=1.0\linewidth,height=0.8\linewidth]{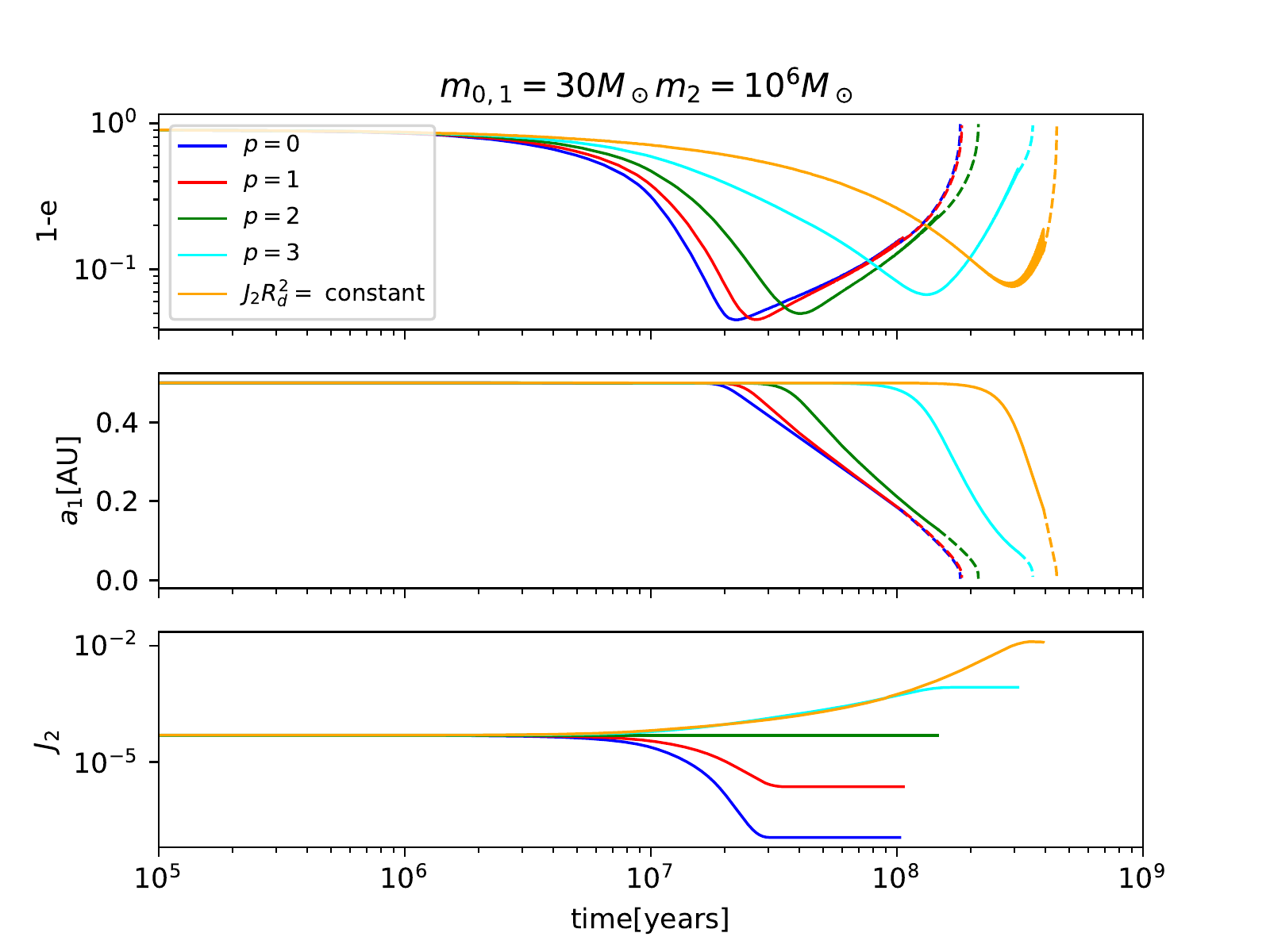}
	\caption{Evolution of binaries with different circum-blackhole disk profiles. Top panel shows eccentricity, middle panel shows semi-major axis and the bottom panel shows $J_2$ as a function of time. Different curves correspond to different disk profiles. We can see that disks with low p values (shallow density profile) allow for higher eccentricity excitation. This is consistent with the fact that low p values cause greater disk disruption which in tern lowers the values of $J_2$ (bottom panel). Consequently, this causes the eccentricity to be excited even further. Merger times are similar which is set by the migration timescale ($\tau_{a_2}=10^8$ years for this figure). It should be noted that the blue curve corresponds to constant disk density and the orange curve corresponds to a constant $J_2R_d^2$ term. We use the following parameters to make these plots: $a_1=0.5 \text{AU}, J_2(0)R^2_d(0)=10^{-5}\text{AU}^2, m_{0,1}=30M_\odot,m_2=10^6M_\odot,e_{1,0}=0.1,a_2=12122.1 \text{AU}$.}
	\label{fig:var_j2_p2}
\end{figure}

\begin{figure}
	\centering
	\includegraphics[width=1.0\linewidth,height=0.75\linewidth]{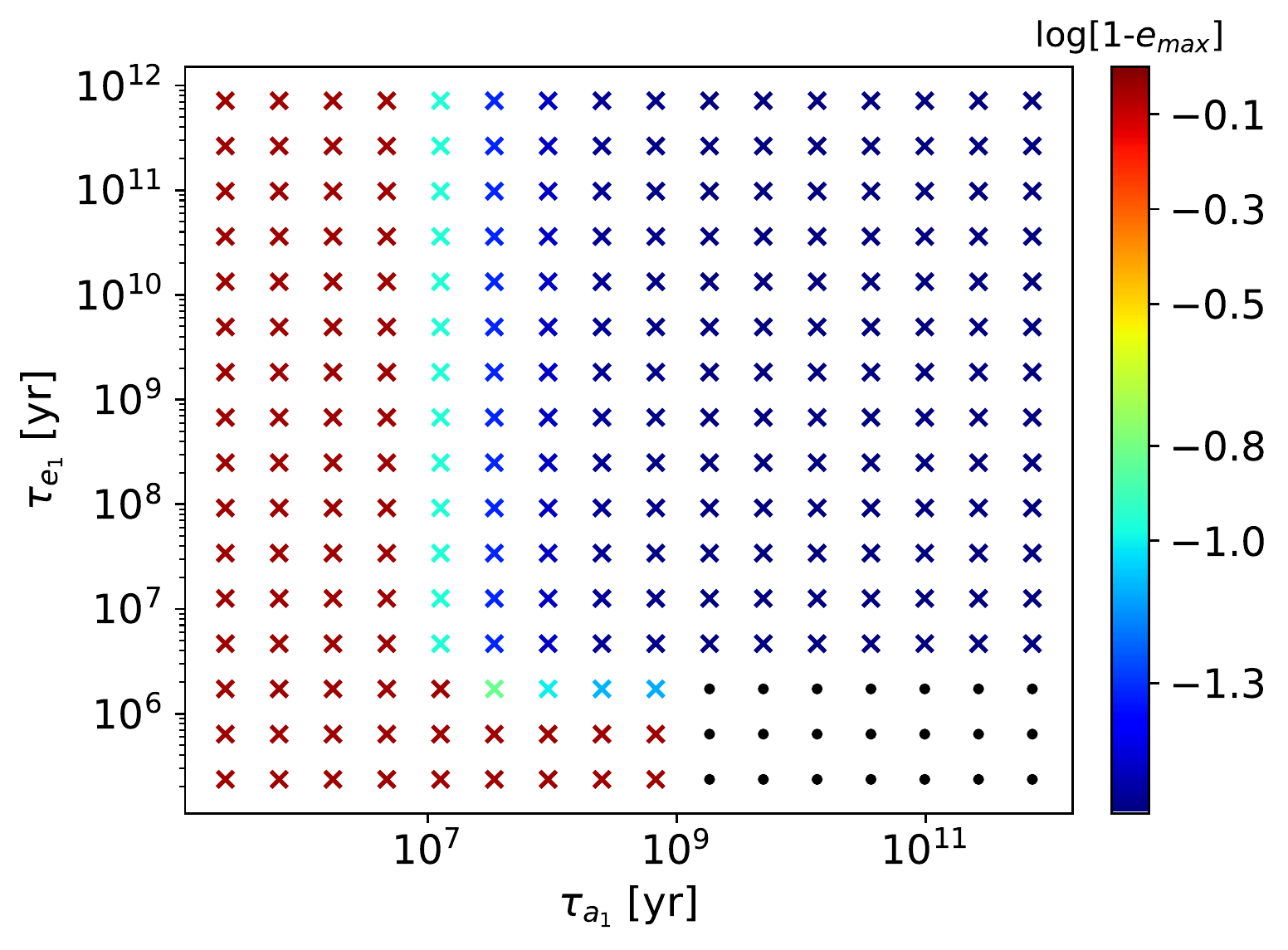}
	
	\caption{Results of ensemble simulations: distribution of maximum eccentricity. The orbital decay timescale ($\tau_{a_1}$) of the binary is shown on the x axis and the circularization timescale ($\tau_{e_1}$) is shown on the y axis. Regions where binaries merge are marked as crosses and where they don't are shown as black circles. Color represents the maximum eccentricity attained in the simulation. When the circularization timescale is long ($> 10^8$ yrs), eccentricity can be excited allowing binaries to merge even when the orbital decay timescale is long. We use the following parameters to get these results: $\tau_{a_2}=10^9 \text{ yrs}, J_2(0)R_d^2(0)=10^{-4} \text{AU}^2 (J_2=4\times10^{-4}),m_2=10^6M_\odot, a_1 = 1 \text{ AU}, m_0=m_1=60 M_\odot$ and initial binary eccentricity $e_{10}=0.01$. We use a variable $J_2$ prescription derived in Eqn. \ref{eqn:varj2} with $p=0$.  For this configuration $\tau_{lib}=2.4\times 10^6$ yrs.}
	\label{fig:en_a1_e1}
\end{figure}

\subsection{Hardening and Circularization of Binaries}
\label{sec:a1e1}
Next, we consider the hardening and circularization of the binary components. As the binary hardens, pericenter distance of the binary decreases causing truncation of the circum-blackhole disk. This causes the reduction of the quadrupole moment of the disc, and the resulting $J_2$ sensitively depends on the disk surface density profile. Thus, we consider different disk density profiles in Fig. \ref{fig:var_j2_p2}. Note that including hardening of the binary alone (decrease in $a_1$), evection resonances do not lead to eccentricity excitation. Thus, we allow the binary to migrate towards the SMBH, on a timescale of $\tau_{a_2}=10^8$ years. Figure \ref{fig:var_j2_p2} shows evolution of equal mass binaries ($m_{0,1}=30 M_\odot$) migrating towards a $10^6 M_\odot$ supermassive blackhole. Different curves represent different disk density profiles (see Eqn. \ref{eqn:varj2}). By comparing different curves we can see that disks with shallow disk density profile (lower p values) experience greater disk dissipation. This causes their eccentricity to be excited to higher values (see top panel). The dashed portion of the trajectory corresponds to when the binary leaves the resonance.

Effects of orbital decay and circularization due the AGN disk are shown in Figure \ref{fig:en_a1_e1} for a constant disk density profile (with $p = 0$). The decay timescale of the semi-major axis and eccentricity are shown on the x-axis and y-axis respectively. The binary itself is migrating on a timescale of $\tau_{a_2}=10^8$ years. Regions of parameter space where the binary merges are marked as crosses and regions where they don't are marked in black circles. The binary starts at an initial semi-major axis of 1 AU. Color shows the maximum eccentricity attained in the simulation. 

We can see that mergers always happen when the orbital decay timescale($\tau_{a_1}$) is short ($<10^9$ years). 
%({\color {green} What is the criterion for mergers to happen, $\tau_{\rm merge} \leq 5$ Gyr?}) 
When it is longer, whether or not binaries merge depends on $\tau_{e_1}$. When $\tau_{e_1}$ is less than 10 million years, binaries are quickly circularized and hence mergers don't happen. On the other hand, when the circularization timescale is longer, binaries can merge due to evection resonances. Due to slower orbital decay rate in this region, the binaries would not have merged if they were not trapped in evection resonances. 

\begin{figure*}
	\centering
	\includegraphics[width=1.0\linewidth,height=0.35\linewidth]{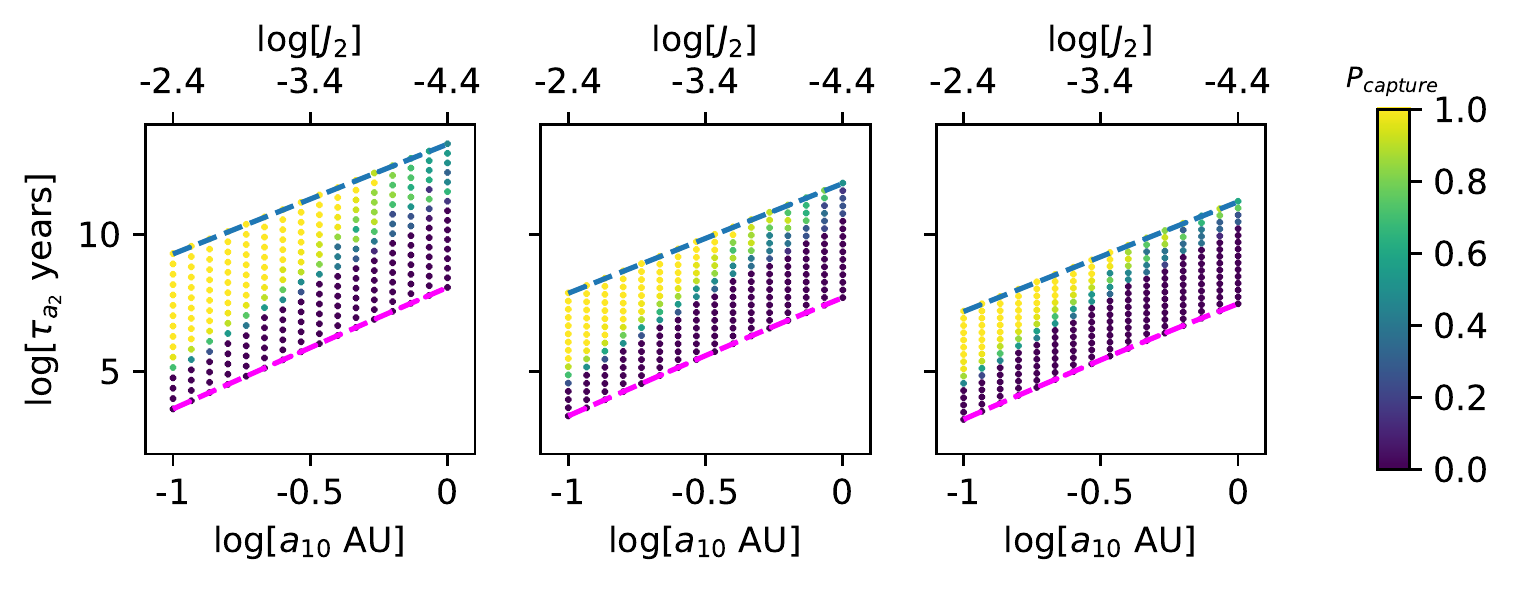}
	
	\caption{Probability of capturing the black-hole binary in evection resonance. The binary migration timescale is shown on the y-axis and the initial semi-major axis of the binary is shown on the x axis.  Color shows the resonance capture probability. The three panels correspond different masses of binaries with $m_0=m_1=$: $20 M_\odot$ (left), $60 M_\odot$ (middle) and $100 M_\odot$ (right). Migration time is sampled between libration time (magenta line) and GR merger time (blue). The initial conditions are same as those used in Figure \ref{fig:en_a2a}. We can see that the binary can always be captured in resonance when the migration timescale is slow ($\tau_{a_2} = 10^8$ years). When the migration timescale is fast ($<10-100 \tau_{lib}$), binaries cannot be captured in evection resonances. The migration timescale corresponding to the boundary between the two regimes increases with the semi-major axis. 
	}
	\label{fig:en_resprob}
\end{figure*}

\subsection{ Resonance Capture Probability}
\label{sec:prob}

So far, we start the binaries in resonance to reduce computational time. However, not all binaries can be captured in resonance depending on the migration rate. In this section, we ran additional simulations to calculate resonance capture probability.

For these simulations we start the binary outside the evection resonance and iterate over different values of the initial phase $\sigma$. Initial semi-major axis of the companion ($a_2$) is chosen such that the binary starts at $\eta=3$.  Since the capture probability  at low eccentricity depends on $d\eta/dt$ \citep{xu_disruption_2016}, we calculate $d\eta/dt$ at the location of resonance (i.e at $\dot{\sigma}(a_1,e_1,a_2)=0$) and migrate the binary such that $d\eta/dt$ is constant.

The results of these simulations are shown in Figure \ref{fig:en_resprob}. Depending on the initial phase and the migration timescale, the binary could be captured in resonance with certainty (when the binary migrates slowly), or with some probabilities. When the migration is fast ($\tau_{a_2}<100\tau_{lib}$), the binary cannot be capture in resonance. 

\section{Conclusions and Discussions}
\label{section:conc}
In this paper we propose an alternative pathway for compact binaries to merge. We show that a binary trapped in evection resonance can have its eccentricity excited allowing its semi-major axis to decay via gravitational radiation. The maximum eccentricity that can be achieved depends on the precession rate of the binary.  Lower the precession rate, higher the eccentricity excitation (Figure \ref{fig:stbpts}). Eccentricity is excited only when the binary sweeps through resonances which happens when the migration timescale is less than libration timescale (Figure \ref{fig:migration}). The libration timescale decreases as the precession rate of the binary increases (Figure \ref{fig:libts}). Hence, merger time is shorter for more massive circum-blackhole disks with faster precession rate, given the eccentricity is excited to the point where gravitational radiation dominates.

We ran an ensemble of simulations to delinate the parameter space where binaries can merge. We find that binaries can merge as long as the migration timescales are greater than 10-100 times the libration timescale. For a typical AGN disk with migration timescales of $10^{7-8}$yrs, the binaries are more likely to merge due to evection resonances if they are more massive ($\sim 100$M$_{\odot}$) or if they are surrounded by a disk ($J_2R_d^2 \sim 10^{-3}$AU$^2$ or $J_2 \sim 10^{-4}$). The merger timescale increases with the semi-major axis (roughly as $a_1^{4.5}$) and decreases with mass. We find that eccentricity excitation due to evection resonance can decrease the merger time by 3-5 orders of magnitude. We also considered how orbital decay and circularization affect binary merger. We find that as long as circularization timescale is long, binary eccentricity can be excited allowing them to merge even when the orbital decay timescale is long.

Multiple studies in literature have explored mechanisms which can harden binaries in gaseous discs (e.g. \cite{grishin2016,Li2022}). For instance, it has been suggested that dynamical friction due to the gas in the AGN disk can reduce the merger time. \cite{bartos_rapid_2017} find that the merger time due to dynamical friction can vary from $10^3$ to $10^6$ years for a 10 $M_\odot$ equal mass binary initially separated at 1 AU. As expected, due to higher gas density the merger time decreases with the semi-major axis of the binary around the SMBH. At $10^{-2}$ pc the merger time is greater than $10^5$ years.  Gravitational torques from the AGN disk can also harden the binary. This is similar to Type II migration discussed above. \cite{baruteau_binaries_2010} find the hardening timescale for this mechanism to be a few million years. Comparing with Figure \ref{fig:libts}, we can see that libration time is less than this limit for $J_2R_d^2 > 10^{-4}$ AU$^2$ for $10 M_\odot$ binaries. For massive binaries, the libration time is much smaller. Hence we expect evection resonance to be important for high mass binaries as well as low mass binaries with massive disks. It is also possible for binaries to soften due to gravitational torques from the disc \citep{yaping2021}. In such a scenario, binaries can still be trapped in evection resonances and have their eccentricity excited \footnote{From Eqns. \ref{eqn:omej2}, \ref{eqn:omegr} we can see that eccentricity an be excited by increasing $a_1$}. \cite{yaping2021} have indeed found in their hydrodynamical simulations, that the eccentricities of blackhole binaries embedded in AGN discs can be excited when $|\eta| \approx 1$. In these simulations, blackhole binaries are subjected to gravitational torques from the disc which causes them to migrate towards the supermassive blackhole and have their semi-major axis increased. In addition, recently it has been reported that under certain conditions, objects in a disc can experience negative dynamical friction which would also increase semi-major axis of a binary \citep{gruzinov2020}. %\li{How does it affect our results? e.g., Thus, we expect that evection resonances could still dominate in the massive disks.} 

In addition to gas in the disk, binaries can interact with other stars/BHs in the disk or the nuclear star cluster surrounding the SMBH which can either harden or ionize the binary depending on the dispersion speed of the star cluster and the binding energy of the binary. \cite{stone_assisted_2017}  study hardening rate due to stellar encounters in AGN disks and find that for $a_{1} <  1$ AU, gas hardening rate is much faster than the stellar hardening rate for $a_2<$ 10 pc. However, they do point out that the stellar hardening could be much higher at larger radii. Hence, for the range of binary separations we consider in this study, stellar encounters are less important than hardening due to gas in the disk. Therefore, evection resonances could play a significant role in enhancing the merger rates. We note that near the completion of our work, we became aware that \citet{Munoz22} have also worked on mergers of black hole binaries due to evection resonances in AGN disks.

\begin{acknowledgments}
The authors thank the referee Evgeni Grishin for helpful comments that substantially improved the quality of the paper. GL and HB are grateful for the support by NASA 80NSSC20K0641 and 80NSSC20K0522. This work used the Hive cluster, which is supported by the National Science Foundation under grant number 1828187.  This research was supported in part through research cyberinfrastrucutre resources and services provided by the Partnership for an Advanced Computing Environment (PACE) at the Georgia Institute of Technology, Atlanta, Georgia, USA.
\end{acknowledgments}

\bibliography{ref.bib}{}
\bibliographystyle{aasjournal}
\end{document}